\documentclass[12pt,final,onecolumn]{IEEEtran}
\usepackage{amsmath}
\usepackage{amsthm}
\usepackage{amsfonts}
\usepackage{amssymb}
\usepackage[utf8]{inputenc}
\DeclareUnicodeCharacter{00A0}{ }
\usepackage{xcolor,mdframed}
\usepackage{graphicx}
\usepackage{booktabs}
\usepackage{graphics}
\usepackage{float}
\usepackage{tikz}
\usetikzlibrary{shapes,snakes,patterns}
\usepackage{epstopdf}
\renewcommand{\vec}[1]{\mathbf{#1}}
\usepackage[justification=centering]{caption}
\usepackage{enumerate} 
\usepackage{cite}
\usepackage{placeins}
\usepackage{algorithm}
\usepackage{hyperref}
\usepackage[noend]{algpseudocode}
\newcommand{\vast}{\bBigg@{4}}
\newcommand{\Vast}{\bBigg@{5}}
\newtheorem{theorem}{Theorem}[]

\newtheorem{lemma}[]{Lemma}
\usepackage{suffix}
\usepackage{hyperref}
\usepackage{mathtools}
\usepackage{amsthm}
\usepackage{enumitem}
\usepackage{xfrac}
\usepackage{relsize}
\theoremstyle{definition}

\usepackage{caption}
\usepackage{subcaption}
\usepackage[normalem]{ulem}
\usepackage{pgfplots}
\pgfplotsset{compat=newest}
\usepackage{xcolor}

\begin{document}
\bstctlcite{IEEEexample:BSTcontrol}
\title{{Optimal Phase Shift Design For Fair Allocation in RIS Aided Uplink Network Using Statistical CSI}
	\author{Athira Subhash, Abla Kammoun, Ahmed Elzanaty, Sheetal Kalyani, Yazan H. Al-Badarneh, and Mohamed-Slim Alouini} \thanks{A. Subhash, and S. Kalyani are with the Department of Electrical Engineering, Indian Institute of Technology, Madras, India. (email:\{ee16d027@smail,skalyani@ee\}.iitm.ac.in).\\A. Elzanaty is with the 5GIC \& 6GIC, Institute for Communication Systems (ICS), University of Surrey, Guildford, GU2 7XH, United Kingdom (e-mail: a.elzanaty@surrey.ac.uk)
\\Y. H. Al-Badarneh is with the Department of Electrical Engineering, The University of Jordan, Amman, 11942 (email: yalbadarneh@ju.edu.jo). \\ A. Kammoun and M.-S. Alouini are with the Computer Electrical and Mathematical Sciences and Engineering (CEMSE) Division, King Abdullah University of Science and Technology (KAUST), Thuwal, Makkah Province, Saudi Arabia. (e-mail:\{abla.kammoun,slim.alouini\}@kaust.edu.sa).}}
\maketitle
\begin{abstract}
Reconfigurable intelligent surfaces (RIS) can be crucial in next-generation communication systems. However, designing the {RIS} phases according to the instantaneous channel state information (CSI) can be challenging in practice due to the short coherent time of the channel. In this regard, we propose a novel algorithm based on the channel statistics of massive multiple input multiple output systems rather than the instantaneous {CSI}. The beamforming at the base station (BS), power allocation of the users, and phase shifts at the RIS elements are optimized to maximize the minimum signal-to-interference and noise ratio (SINR), guaranteeing fair operation among various users. In particular, we design the RIS phases by leveraging the asymptotic deterministic equivalent of the minimum {SINR} that depends only on the channel statistics. This significantly reduces the computational complexity and the amount of controlling data between the {BS} and {RIS} for updating the phases. This setup is also useful for electromagnetic fields (EMF)-aware systems with constraints on the maximum user's exposure to EMF. The numerical results show that the proposed algorithms achieve more than $100 \%$ gain in terms of minimum SINR, compared to a system with random RIS phase shifts, when $40$ RIS elements, $20$ antennas at the BS and $10$ users, are considered.
 \end{abstract}

\section{Introduction}
The idea of reconfigurable intelligent surfaces (RIS) capable of reconfiguring the wireless propagation environment has recently attracted much research attention \cite{basar2019wireless,ozdogan2019intelligent}. RISs, also widely referred to as intelligent reflecting surfaces (IRS), are surfaces made of metamaterials capable of modifying the properties of electromagnetic radiation impinging them according to some control mechanisms \cite{di2020smart}. The modified properties may include the amplitude, phase, frequency or polarization of the impinging wave. The most commonly studied RIS model is the one changing the phase of the incident wave \cite{basar2019wireless}. RIS usually comprises an array of reflecting elements, each capable of individually modifying the properties of the incident wave and the associated control mechanism. Note that such a structure can modify the wireless propagation environment and hence can be used to achieve beam-steering and focusing \cite{di2020smart}. Therefore RIS is a potential enabler for the concept of smart radio environments. Conventionally, improvements in wireless system performance were achieved by improving the transmitter and receiver design to combat the detrimental effects of the wireless propagation channel. Smart radio environments aim to introduce controlled changes in the wireless channel to improve signal propagation \cite{di2020smart,liaskos2018new}, and a RIS proves to be one enabler for the same. 
\par Several works in the literature have studied the prospects of potential improvements in the wireless system performance by deploying RIS to improve cell coverage, compensate for the absence of a line of sight (LOS) link, enhance the physical layer security, in millimetre (mm) wave channels and several other applications \cite{pan2020multicell,perovic2020channel,yu2019enabling,gopi2020intelligent}. The authors of \cite{perovic2020channel} study the channel capacity optimization of single-stream transmission using RISs in a mmWave system without any LOS path. They propose a scheme to optimize the RIS phase shifts and the transmit phase precoder to maximize the capacity. The authors of \cite{guo2020weighted} maximize the weighted sum rate of the users by jointly designing the beamforming (BF) at the access point and the phase shifts of the RIS elements in a downlink multiple input single output (MISO) systems. The authors of \cite{yu2019miso} propose a joint optimization strategy to select the optimal beamformer at the access point (AP) and the RIS phase shifts to maximize the
spectral efficiency (SE) in the downlink of a single-user MISO system. A deep reinforcement-based resource allocation is proposed for a single-user MISO system by the authors of \cite{feng2020deep}. The downlink multi-user MISO systems are considered by the authors of \cite{alwazani2020intelligent,abeywickrama2020intelligent}. 
An algorithm to find the optimal BF strategy and phase shift design that minimizes transmit power at the AP, subject to constraints on the signal-to-interference and noise ratio (SINR) of each user, is proposed in \cite{abeywickrama2020intelligent}. The authors of \cite{luo2021spatial} study the resource allocation in a RIS-aided spatially modulated multiple input multiple output (MIMO) uplink network. They propose an algorithm for jointly optimizing the users' transmit power and the RIS phase shifts to minimize the system symbol error. A fair resource allocation strategy to maximize the minimum SINR is discussed in \cite{alwazani2020intelligent}. They also study the susceptibility of the system to channel estimation errors. 
\par In all of the above works, the availability of exact channel state information (CSI) at all the nodes is assumed. Note that this assumption might not be practical in all scenarios. Moreover, given the time-varying nature of the wireless channel, resource allocation strategies using the exact CSI will be valid only over the coherence time and must be updated frequently. This would, in turn, increase the computational complexity and the feedback overhead, especially in updating the RIS phase shifts. Using statistical CSI can significantly reduce the channel estimation overhead, and computational complexity in such scenarios \cite{zhi2021statistical}. Several authors have explored the prospects of using statistical CSI for resource allocation problems in RIS-aided systems. \par A single-user MIMO system assisted by a RIS is studied by the authors of \cite{wang2021joint}. Assuming that the statistical CSI is known at both the transmitter and the RIS, they propose a resource allocation scheme to maximize the ergodic SE. An upper bound on the SE is derived and utilized to develop the resource allocation strategy. The authors of \cite{you2021reconfigurable} propose a strategy for designing the transmit covariance matrices of the users and the RIS phase shift matrix to maximize the system's global energy efficiency (GEE) with partial CSI assuming the correlated Rayleigh model. Using results from random matrix theory (RMT), asymptotic deterministic equivalents of the GEE are first derived and then used to solve the optimization problem. The authors of \cite{shi2021outage} study the outage performance of RIS-aided MIMO communications over
cascaded Rayleigh fading channels. A genetic algorithm is used to find the optimal phase shift design that minimizes the asymptotic outage probability. The outage probability at the destination of a RIS-assisted communication system in a $\kappa-\mu$ fading environment is studied by the authors of \cite{charishma2021outage}. 

The downlink of a multi-user MISO system is studied by the authors of \cite{kammoun2020asymptotic}. They solve the problem of determining the optimal linear precoder, the power allocation matrix at the BS and the RIS phase shifts matrix to maximize the system's minimum SINR. Assuming a full-rank BS-to-RIS channel matrix results from RMT are used to derive deterministic approximations for the minimum user SINR. This is, in turn, used to propose the optimal phase shift design algorithm. Similarly, the downlink of a multi-user MIMO system supported by a RIS is studied by the authors of \cite{abrardo2020intelligent}. They propose an optimization algorithm to configure the RISs to maximize the network sum rate. The proposed solution depends only on the distribution of the users' locations and the distribution of the multipath channel, hence avoiding the need for frequently reconfiguring the RIS. A brief overview of the differences in the models studied in a few of the key literature discussed above is presented in Table \ref{literature_summary}. 
\par To the best of our knowledge, the interesting scenario of fair resource allocation in a RIS-aided uplink network with constraints on the per-user transmit power has not been studied in literature so far. Such a fair resource allocation scheme will be of interest in several multi-user scenarios. Moreover, a resource allocation strategy that does not need to be repeated for every channel realization will be of utility in any practical scenario. Since the RIS phase shift design is usually evaluated at some access point and later fed back to the RIS controller, a RIS phase shift design depending on the exact channel statistics, will lead to a large feedback overhead. Hence, it is of particular interest to explore a solution for the optimal RIS phase, which will depend only on the channel statistics. Such a solution will be valid for a longer time and considerably reduce the feedback overheads. 
\par In this regard, we consider the uplink communication of a multi-user single input multiple output (SIMO) system assisted by a multi-element RIS with correlated elements. We study the joint optimization of the receive BF vectors at the BS, transmit power at the users and the phase shift vector at the RIS that maximizes the minimum SINR. An alternating optimization strategy is proposed to solve the joint resource allocation problem. The BF vector design and the transmit power allocation assume the availability of the exact CSI and are implemented using the results from \cite{emfrisinst}. To simplify the computational complexity and avoid frequent feedback, the phase shifts at the RIS are designed using the statistical CSI. More elaborately, we use tools from RMT to derive the asymptotic deterministic equivalent of the minimum user SINR for the large system dimensions. Later, the phase shift optimization is performed using this deterministic equivalent. We also demonstrate the utility of the proposed results for the joint resource allocation of RIS-aided communication systems with constraints in the maximum electromagnetic field (EMF) exposure. 

 Recently, the EMF exposure of the users has been a concern in the fifth-generation communication systems \cite{chiaraviglio2021health}. There has been an increase in the number of users concerned about the effects of continued exposure to EMF, and there are many works studying this problem in detail {\cite{chiaraviglio2021health,sambo2014survey,wang2011evaluation,Elzanaty_635716,zappone2022energy,phan2022creating}}. The authors of \cite{chiaraviglio2021health} discuss the health effects of EMF exposure, metrics to characterize the exposure and a review of the different approaches to reduce the exposure in 5G systems. The authors of \cite{wang2011evaluation} propose a technique for evaluating and optimizing a metric called the specific absorption rate (SAR) for multi-antenna wireless systems. The SAR, measured in Watts per kilogramme of body weight (W/kg), quantifies the EMF exposure  in the near-field region. It measures the rate of energy absorption by the body when exposed to EM radiation. EMF exposure from mobile phones is mostly in the near-field, so SAR is used to evaluate such exposure over a certain period of time.

 The problem of optimal resource allocation that minimizes the EMF exposure of users while maintaining the quality of service (QoS) requirements has received much attention recently \cite{ibraiwish2021emf,javedemf,heliot2021minimizing}. The authors in \cite{javedemf} propose a  scheme to minimize human exposure to EMF radiations while achieving the target throughput using probabilistic shaping. The authors of \cite{heliot2021minimizing} propose an optimization strategy to minimize the average exposure dose of users, considering both the system's QoS and transmit power constraints. A novel architecture that exploits a RIS to minimize the EMF exposure is firstly proposed in \cite{ibraiwish2021emf} were the RIS phases and the transmit power of users are optimized.   The joint optimization of the RIS phase shifts,  transmit BF,  linear receive filter, and  transmit power to maximize the energy efficiency in a RIS-aided network subject to constraints on the maximum user EMF is studied in \cite{zappone2022energy}. The feasibility of a {'Reduced EMF Exposure Area'} using RIS-aided networks is then studied in \cite{phan2022creating}.

 \begin{table*}[t!]
 	\centering
 	\caption{{Key literature studying RIS phase optimization.}}
 	\begin{tabular}{|l|l|l|l|l|}
 		\hline
 		Reference & \begin{tabular}[c]{@{}l@{}}Instantaneous CSI/\\ Statistical CSI\end{tabular} & Objective &  \begin{tabular}[c]{@{}l@{}}Uplink/ \\ Downlink\end{tabular} &  \begin{tabular}[c]{@{}l@{}}Antenna\\  model\end{tabular}            \\  \hline
 		\cite{yu2019miso} & Instantaneous & SE & Downlink & MISO   \\
 		\hline 
 		\cite{feng2020deep} & Instantaneous & Received SNR & Downlink & MISO   \\
 		\hline 
 		\cite{abeywickrama2020intelligent} & Instantaneous & Achievable rate & Downlink & MISO   \\
 		\hline 
 		\cite{zhi2021statistical} & Statistical & Ergodic rate & Uplink & MIMO  \\
 		\hline 
 		\cite{wang2021joint} & Statistics & Ergodic spectral efficiency & Uplink & MIMO  \\
 		\hline 
 		\cite{you2021reconfigurable} & Partial & Energy efficiency &Uplink & MIMO  \\
 		\hline 
 		\cite{kammoun2020asymptotic} & Instantaneous  & Minimum SINR & Downlink & MISO  \\
 		\hline
 		\cite{luo2021spatial} & Instantaneous & Symbol error rate & Uplink & MIMO \\
 		\hline 
 		\cite{emfrisinst} & Instantaneous & Minimum SINR & Uplink & SIMO \\
 		\hline
 		This work & Statistical & Minimum SINR & Uplink & SIMO \\
 		\hline
 	\end{tabular}
 	\label{literature_summary}
 \end{table*}

 Given that our resource allocation scheme maximizes the minimum user SINR, we must ensure that the resulting solution will not result in users experiencing more EMF radiation than the stipulated thresholds. Hence, we demonstrate how the problem can be studied with constraints imposed on the maximum EMF exposure of each user. The main contributions of this paper can be summarized as follows.
\begin{itemize}
\item We formulate the problem of jointly optimizing the receive BF vectors at the BS, transmit power allocation at the users, and the RIS phase shift design in a RIS-aided multi-user system with multiple antennas at the BS. We discuss solutions for the BF and power allocation, assuming the availability of exact CSI.
\item We then study the asymptotic statistics of the minimum SINR for this BF choice at the BS. We use tools from RMT to derive the asymptotic deterministic equivalent of received SINR.
\item A simple projected gradient descent-based solution is proposed to identify the optimal phase shift choice at the RIS that maximizes the asymptotic minimum user SINR.
\item This solution has less computational complexity and depends on large-scale channel gains. Thus, the proposed scheme also avoids the need for frequent reconfiguration of the RIS with costly feedback.
\end{itemize}    
    
\par The rest of the paper is organized as follows. Section \ref{sys_model} presents the system model, and Section \ref{opti_prob} explains the proposed optimization problem and the corresponding solution. Section \ref{application} discusses one application for the proposed results, and Section \ref{simu} demonstrates the results of simulation experiments. Finally, we conclude the work in Section \ref{conclusion}
 
\subsection{Notations}
The notations used in this paper are as follows: $\mathbb{C}^{p\times q}$ represents a complex matrix of dimensions $p$ by $q$, $\mathcal{CN}(\mu,\sigma^2)$ represents the complex normal random variable (RV) with mean $\mu$ and variance $\sigma^2$, $\text{diag}\left( a_1,\cdots,a_N\right)$ represents a diagonal matrix with diagonal elements $a_1,\cdots,a_N$, and $||\vec{a}||$ represents the $\ell_2$ norm of the vector $\vec{a}$. $\text{unif}[a,b]$ represents the uniform distribution over the support $[a,b]$. $\left[ A\right]_{i,j}$ represents the $(i,j)$-th element of the matrix $A$. Matrices are denoted by boldface uppercase letters, vectors are denoted by boldface lowercase letters and scalars are represented using lowercase letters. 

\section{System Model}\label{sys_model}
\begin{figure*}[t!]
	\centering
	\tikzset{every picture/.style={line width=0.75pt}} 
	\resizebox{100mm}{50mm}{       
	\begin{tikzpicture}[x=0.75pt,y=0.75pt,yscale=-1,xscale=1]
		\draw    (104.33,182) -- (138.33,281) ;
		\draw    (104.33,182) -- (72.33,281) ;
		\draw    (94,215) -- (119.33,281) ;
		\draw    (83.33,247) -- (95.33,280) ;
		\draw    (116.33,219) -- (95.33,280) ;
		\draw    (129.33,253) -- (119.33,281) ;
		\draw    (52.33,182) -- (156.33,182) ;
		\draw    (153.33,147) -- (153.33,182) ;
		\draw   (142.33,164) .. controls (148.36,157.34) and (151.97,150.67) .. (153.17,144) .. controls (154.37,150.67) and (157.98,157.34) .. (164,164) ;
		\draw    (53.33,148) -- (53.33,183) ;
		\draw   (42.33,165) .. controls (48.36,158.34) and (51.97,151.67) .. (53.17,145) .. controls (54.37,151.67) and (57.98,158.34) .. (64,165) ;
		\draw    (86.33,148) -- (86.33,183) ;
		\draw   (75.33,165) .. controls (81.36,158.34) and (84.97,151.67) .. (86.17,145) .. controls (87.37,151.67) and (90.98,158.34) .. (97,165) ;
		\draw  [fill={rgb, 255:red, 0; green, 0; blue, 0 }  ,fill opacity=1 ] (108.33,166.67) .. controls (108.33,165.38) and (109.38,164.33) .. (110.67,164.33) .. controls (111.96,164.33) and (113,165.38) .. (113,166.67) .. controls (113,167.96) and (111.96,169) .. (110.67,169) .. controls (109.38,169) and (108.33,167.96) .. (108.33,166.67) -- cycle ;
		\draw  [fill={rgb, 255:red, 0; green, 0; blue, 0 }  ,fill opacity=1 ] (128.33,166.67) .. controls (128.33,165.38) and (129.38,164.33) .. (130.67,164.33) .. controls (131.96,164.33) and (133,165.38) .. (133,166.67) .. controls (133,167.96) and (131.96,169) .. (130.67,169) .. controls (129.38,169) and (128.33,167.96) .. (128.33,166.67) -- cycle ;
		\draw  [fill={rgb, 255:red, 242; green, 237; blue, 237 }  ,fill opacity=1 ] (62.33,281) -- (150.33,281) -- (150.33,297) -- (62.33,297) -- cycle ;
		\draw  [fill={rgb, 255:red, 239; green, 232; blue, 232 }  ,fill opacity=1 ] (204,5) -- (369.33,5) -- (369.33,69) -- (204,69) -- cycle ;
		\draw  [fill={rgb, 255:red, 239; green, 232; blue, 232 }  ,fill opacity=1 ] (204,5) -- (227.62,5) -- (227.62,21) -- (204,21) -- cycle ;
		\draw  [fill={rgb, 255:red, 239; green, 232; blue, 232 }  ,fill opacity=1 ] (204,37) -- (227.62,37) -- (227.62,53) -- (204,53) -- cycle ;
		\draw  [fill={rgb, 255:red, 239; green, 232; blue, 232 }  ,fill opacity=1 ] (204,21) -- (227.62,21) -- (227.62,37) -- (204,37) -- cycle ;
		\draw  [fill={rgb, 255:red, 239; green, 232; blue, 232 }  ,fill opacity=1 ] (204,53) -- (227.62,53) -- (227.62,69) -- (204,69) -- cycle ;
		\draw  [fill={rgb, 255:red, 239; green, 232; blue, 232 }  ,fill opacity=1 ] (227.62,5) -- (251.24,5) -- (251.24,21) -- (227.62,21) -- cycle ;
		\draw  [fill={rgb, 255:red, 239; green, 232; blue, 232 }  ,fill opacity=1 ] (227.62,37) -- (251.24,37) -- (251.24,53) -- (227.62,53) -- cycle ;
		\draw  [fill={rgb, 255:red, 239; green, 232; blue, 232 }  ,fill opacity=1 ] (227.62,21) -- (251.24,21) -- (251.24,37) -- (227.62,37) -- cycle ;
		\draw  [fill={rgb, 255:red, 239; green, 232; blue, 232 }  ,fill opacity=1 ] (227.62,53) -- (251.24,53) -- (251.24,69) -- (227.62,69) -- cycle ;
		\draw  [fill={rgb, 255:red, 239; green, 232; blue, 232 }  ,fill opacity=1 ] (251.24,5) -- (274.86,5) -- (274.86,21) -- (251.24,21) -- cycle ;
		\draw  [fill={rgb, 255:red, 239; green, 232; blue, 232 }  ,fill opacity=1 ] (251.24,37) -- (274.86,37) -- (274.86,53) -- (251.24,53) -- cycle ;
		\draw  [fill={rgb, 255:red, 239; green, 232; blue, 232 }  ,fill opacity=1 ] (251.24,21) -- (274.86,21) -- (274.86,37) -- (251.24,37) -- cycle ;
		\draw  [fill={rgb, 255:red, 239; green, 232; blue, 232 }  ,fill opacity=1 ] (251.24,53) -- (274.86,53) -- (274.86,69) -- (251.24,69) -- cycle ;
		\draw  [fill={rgb, 255:red, 239; green, 232; blue, 232 }  ,fill opacity=1 ] (274.86,5) -- (298.48,5) -- (298.48,21) -- (274.86,21) -- cycle ;
		\draw  [fill={rgb, 255:red, 239; green, 232; blue, 232 }  ,fill opacity=1 ] (274.86,37) -- (298.48,37) -- (298.48,53) -- (274.86,53) -- cycle ;
		\draw  [fill={rgb, 255:red, 239; green, 232; blue, 232 }  ,fill opacity=1 ] (274.86,21) -- (298.48,21) -- (298.48,37) -- (274.86,37) -- cycle ;
		\draw  [fill={rgb, 255:red, 239; green, 232; blue, 232 }  ,fill opacity=1 ] (274.86,53) -- (298.48,53) -- (298.48,69) -- (274.86,69) -- cycle ;
		\draw  [fill={rgb, 255:red, 239; green, 232; blue, 232 }  ,fill opacity=1 ] (298.48,5) -- (322.1,5) -- (322.1,21) -- (298.48,21) -- cycle ;
		\draw  [fill={rgb, 255:red, 239; green, 232; blue, 232 }  ,fill opacity=1 ] (298.48,37) -- (322.1,37) -- (322.1,53) -- (298.48,53) -- cycle ;
		\draw  [fill={rgb, 255:red, 239; green, 232; blue, 232 }  ,fill opacity=1 ] (298.48,21) -- (322.1,21) -- (322.1,37) -- (298.48,37) -- cycle ;
		\draw  [fill={rgb, 255:red, 239; green, 232; blue, 232 }  ,fill opacity=1 ] (298.48,53) -- (322.1,53) -- (322.1,69) -- (298.48,69) -- cycle ;
		\draw  [fill={rgb, 255:red, 239; green, 232; blue, 232 }  ,fill opacity=1 ] (322.1,5) -- (345.71,5) -- (345.71,21) -- (322.1,21) -- cycle ;
		\draw  [fill={rgb, 255:red, 239; green, 232; blue, 232 }  ,fill opacity=1 ] (322.1,37) -- (345.71,37) -- (345.71,53) -- (322.1,53) -- cycle ;
		\draw  [fill={rgb, 255:red, 239; green, 232; blue, 232 }  ,fill opacity=1 ] (322.1,21) -- (345.71,21) -- (345.71,37) -- (322.1,37) -- cycle ;
		\draw  [fill={rgb, 255:red, 239; green, 232; blue, 232 }  ,fill opacity=1 ] (322.1,53) -- (345.71,53) -- (345.71,69) -- (322.1,69) -- cycle ;
		\draw  [fill={rgb, 255:red, 239; green, 232; blue, 232 }  ,fill opacity=1 ] (345.71,5) -- (369.33,5) -- (369.33,21) -- (345.71,21) -- cycle ;
		\draw  [fill={rgb, 255:red, 239; green, 232; blue, 232 }  ,fill opacity=1 ] (345.71,37) -- (369.33,37) -- (369.33,53) -- (345.71,53) -- cycle ;
		\draw  [fill={rgb, 255:red, 239; green, 232; blue, 232 }  ,fill opacity=1 ] (345.71,21) -- (369.33,21) -- (369.33,37) -- (345.71,37) -- cycle ;
		\draw  [fill={rgb, 255:red, 239; green, 232; blue, 232 }  ,fill opacity=1 ] (345.71,53) -- (369.33,53) -- (369.33,69) -- (345.71,69) -- cycle ;
		\draw  [fill={rgb, 255:red, 0; green, 0; blue, 0 }  ,fill opacity=1 ] (545.83,127.2) .. controls (545.83,125.82) and (546.95,124.7) .. (548.33,124.7) -- (548.33,124.7) .. controls (549.71,124.7) and (550.83,125.82) .. (550.83,127.2) -- (550.83,138) .. controls (550.83,139.38) and (549.71,140.5) .. (548.33,140.5) -- (548.33,140.5) .. controls (546.95,140.5) and (545.83,139.38) .. (545.83,138) -- cycle ;
		\draw  [fill={rgb, 255:red, 222; green, 217; blue, 217 }  ,fill opacity=1 ] (557.74,146.7) .. controls (557.74,140.35) and (552.6,135.2) .. (546.24,135.2) -- (546.24,135.2) .. controls (539.89,135.2) and (534.74,140.35) .. (534.74,146.7) -- (534.74,161.5) .. controls (534.74,167.85) and (539.89,173) .. (546.24,173) -- (546.24,173) .. controls (552.6,173) and (557.74,167.85) .. (557.74,161.5) -- cycle ;
		\draw   (538,140.17) -- (553.33,140.17) -- (553.33,150.17) -- (538,150.17) -- cycle ;
		\draw  [fill={rgb, 255:red, 62; green, 61; blue, 61 }  ,fill opacity=1 ] (538.07,153.7) .. controls (538.07,152.47) and (539.04,151.47) .. (540.24,151.47) .. controls (541.45,151.47) and (542.42,152.47) .. (542.42,153.7) .. controls (542.42,154.93) and (541.45,155.93) .. (540.24,155.93) .. controls (539.04,155.93) and (538.07,154.93) .. (538.07,153.7) -- cycle ;
		\draw  [fill={rgb, 255:red, 62; green, 61; blue, 61 }  ,fill opacity=1 ] (544.07,153.7) .. controls (544.07,152.47) and (545.04,151.47) .. (546.24,151.47) .. controls (547.45,151.47) and (548.42,152.47) .. (548.42,153.7) .. controls (548.42,154.93) and (547.45,155.93) .. (546.24,155.93) .. controls (545.04,155.93) and (544.07,154.93) .. (544.07,153.7) -- cycle ;
		\draw  [fill={rgb, 255:red, 62; green, 61; blue, 61 }  ,fill opacity=1 ] (550.07,153.7) .. controls (550.07,152.47) and (551.04,151.47) .. (552.24,151.47) .. controls (553.45,151.47) and (554.42,152.47) .. (554.42,153.7) .. controls (554.42,154.93) and (553.45,155.93) .. (552.24,155.93) .. controls (551.04,155.93) and (550.07,154.93) .. (550.07,153.7) -- cycle ;
		\draw  [fill={rgb, 255:red, 62; green, 61; blue, 61 }  ,fill opacity=1 ] (538.07,159.7) .. controls (538.07,158.47) and (539.04,157.47) .. (540.24,157.47) .. controls (541.45,157.47) and (542.42,158.47) .. (542.42,159.7) .. controls (542.42,160.93) and (541.45,161.93) .. (540.24,161.93) .. controls (539.04,161.93) and (538.07,160.93) .. (538.07,159.7) -- cycle ;
		\draw  [fill={rgb, 255:red, 62; green, 61; blue, 61 }  ,fill opacity=1 ] (544.07,159.7) .. controls (544.07,158.47) and (545.04,157.47) .. (546.24,157.47) .. controls (547.45,157.47) and (548.42,158.47) .. (548.42,159.7) .. controls (548.42,160.93) and (547.45,161.93) .. (546.24,161.93) .. controls (545.04,161.93) and (544.07,160.93) .. (544.07,159.7) -- cycle ;
		\draw  [fill={rgb, 255:red, 62; green, 61; blue, 61 }  ,fill opacity=1 ] (550.07,159.7) .. controls (550.07,158.47) and (551.04,157.47) .. (552.24,157.47) .. controls (553.45,157.47) and (554.42,158.47) .. (554.42,159.7) .. controls (554.42,160.93) and (553.45,161.93) .. (552.24,161.93) .. controls (551.04,161.93) and (550.07,160.93) .. (550.07,159.7) -- cycle ;
		\draw  [fill={rgb, 255:red, 62; green, 61; blue, 61 }  ,fill opacity=1 ] (538.07,165.7) .. controls (538.07,164.47) and (539.04,163.47) .. (540.24,163.47) .. controls (541.45,163.47) and (542.42,164.47) .. (542.42,165.7) .. controls (542.42,166.93) and (541.45,167.93) .. (540.24,167.93) .. controls (539.04,167.93) and (538.07,166.93) .. (538.07,165.7) -- cycle ;
		\draw  [fill={rgb, 255:red, 62; green, 61; blue, 61 }  ,fill opacity=1 ] (544.07,165.7) .. controls (544.07,164.47) and (545.04,163.47) .. (546.24,163.47) .. controls (547.45,163.47) and (548.42,164.47) .. (548.42,165.7) .. controls (548.42,166.93) and (547.45,167.93) .. (546.24,167.93) .. controls (545.04,167.93) and (544.07,166.93) .. (544.07,165.7) -- cycle ;
		\draw  [fill={rgb, 255:red, 62; green, 61; blue, 61 }  ,fill opacity=1 ] (550.07,165.7) .. controls (550.07,164.47) and (551.04,163.47) .. (552.24,163.47) .. controls (553.45,163.47) and (554.42,164.47) .. (554.42,165.7) .. controls (554.42,166.93) and (553.45,167.93) .. (552.24,167.93) .. controls (551.04,167.93) and (550.07,166.93) .. (550.07,165.7) -- cycle ;
		
		\draw  [fill={rgb, 255:red, 0; green, 0; blue, 0 }  ,fill opacity=1 ] (473.83,101.2) .. controls (473.83,99.82) and (474.95,98.7) .. (476.33,98.7) -- (476.33,98.7) .. controls (477.71,98.7) and (478.83,99.82) .. (478.83,101.2) -- (478.83,112) .. controls (478.83,113.38) and (477.71,114.5) .. (476.33,114.5) -- (476.33,114.5) .. controls (474.95,114.5) and (473.83,113.38) .. (473.83,112) -- cycle ;
		\draw  [fill={rgb, 255:red, 222; green, 217; blue, 217 }  ,fill opacity=1 ] (485.74,120.7) .. controls (485.74,114.35) and (480.6,109.2) .. (474.24,109.2) -- (474.24,109.2) .. controls (467.89,109.2) and (462.74,114.35) .. (462.74,120.7) -- (462.74,135.5) .. controls (462.74,141.85) and (467.89,147) .. (474.24,147) -- (474.24,147) .. controls (480.6,147) and (485.74,141.85) .. (485.74,135.5) -- cycle ;
		\draw   (466,114.17) -- (481.33,114.17) -- (481.33,124.17) -- (466,124.17) -- cycle ;
		\draw  [fill={rgb, 255:red, 62; green, 61; blue, 61 }  ,fill opacity=1 ] (466.07,127.7) .. controls (466.07,126.47) and (467.04,125.47) .. (468.24,125.47) .. controls (469.45,125.47) and (470.42,126.47) .. (470.42,127.7) .. controls (470.42,128.93) and (469.45,129.93) .. (468.24,129.93) .. controls (467.04,129.93) and (466.07,128.93) .. (466.07,127.7) -- cycle ;
		\draw  [fill={rgb, 255:red, 62; green, 61; blue, 61 }  ,fill opacity=1 ] (472.07,127.7) .. controls (472.07,126.47) and (473.04,125.47) .. (474.24,125.47) .. controls (475.45,125.47) and (476.42,126.47) .. (476.42,127.7) .. controls (476.42,128.93) and (475.45,129.93) .. (474.24,129.93) .. controls (473.04,129.93) and (472.07,128.93) .. (472.07,127.7) -- cycle ;
		\draw  [fill={rgb, 255:red, 62; green, 61; blue, 61 }  ,fill opacity=1 ] (478.07,127.7) .. controls (478.07,126.47) and (479.04,125.47) .. (480.24,125.47) .. controls (481.45,125.47) and (482.42,126.47) .. (482.42,127.7) .. controls (482.42,128.93) and (481.45,129.93) .. (480.24,129.93) .. controls (479.04,129.93) and (478.07,128.93) .. (478.07,127.7) -- cycle ;
		\draw  [fill={rgb, 255:red, 62; green, 61; blue, 61 }  ,fill opacity=1 ] (466.07,133.7) .. controls (466.07,132.47) and (467.04,131.47) .. (468.24,131.47) .. controls (469.45,131.47) and (470.42,132.47) .. (470.42,133.7) .. controls (470.42,134.93) and (469.45,135.93) .. (468.24,135.93) .. controls (467.04,135.93) and (466.07,134.93) .. (466.07,133.7) -- cycle ;
		\draw  [fill={rgb, 255:red, 62; green, 61; blue, 61 }  ,fill opacity=1 ] (472.07,133.7) .. controls (472.07,132.47) and (473.04,131.47) .. (474.24,131.47) .. controls (475.45,131.47) and (476.42,132.47) .. (476.42,133.7) .. controls (476.42,134.93) and (475.45,135.93) .. (474.24,135.93) .. controls (473.04,135.93) and (472.07,134.93) .. (472.07,133.7) -- cycle ;
		\draw  [fill={rgb, 255:red, 62; green, 61; blue, 61 }  ,fill opacity=1 ] (478.07,133.7) .. controls (478.07,132.47) and (479.04,131.47) .. (480.24,131.47) .. controls (481.45,131.47) and (482.42,132.47) .. (482.42,133.7) .. controls (482.42,134.93) and (481.45,135.93) .. (480.24,135.93) .. controls (479.04,135.93) and (478.07,134.93) .. (478.07,133.7) -- cycle ;
		\draw  [fill={rgb, 255:red, 62; green, 61; blue, 61 }  ,fill opacity=1 ] (466.07,139.7) .. controls (466.07,138.47) and (467.04,137.47) .. (468.24,137.47) .. controls (469.45,137.47) and (470.42,138.47) .. (470.42,139.7) .. controls (470.42,140.93) and (469.45,141.93) .. (468.24,141.93) .. controls (467.04,141.93) and (466.07,140.93) .. (466.07,139.7) -- cycle ;
		\draw  [fill={rgb, 255:red, 62; green, 61; blue, 61 }  ,fill opacity=1 ] (472.07,139.7) .. controls (472.07,138.47) and (473.04,137.47) .. (474.24,137.47) .. controls (475.45,137.47) and (476.42,138.47) .. (476.42,139.7) .. controls (476.42,140.93) and (475.45,141.93) .. (474.24,141.93) .. controls (473.04,141.93) and (472.07,140.93) .. (472.07,139.7) -- cycle ;
		\draw  [fill={rgb, 255:red, 62; green, 61; blue, 61 }  ,fill opacity=1 ] (478.07,139.7) .. controls (478.07,138.47) and (479.04,137.47) .. (480.24,137.47) .. controls (481.45,137.47) and (482.42,138.47) .. (482.42,139.7) .. controls (482.42,140.93) and (481.45,141.93) .. (480.24,141.93) .. controls (479.04,141.93) and (478.07,140.93) .. (478.07,139.7) -- cycle ;
		
		\draw  [fill={rgb, 255:red, 0; green, 0; blue, 0 }  ,fill opacity=1 ] (355.83,230.2) .. controls (355.83,228.82) and (356.95,227.7) .. (358.33,227.7) -- (358.33,227.7) .. controls (359.71,227.7) and (360.83,228.82) .. (360.83,230.2) -- (360.83,241) .. controls (360.83,242.38) and (359.71,243.5) .. (358.33,243.5) -- (358.33,243.5) .. controls (356.95,243.5) and (355.83,242.38) .. (355.83,241) -- cycle ;
		\draw  [fill={rgb, 255:red, 222; green, 217; blue, 217 }  ,fill opacity=1 ] (367.74,249.7) .. controls (367.74,243.35) and (362.6,238.2) .. (356.24,238.2) -- (356.24,238.2) .. controls (349.89,238.2) and (344.74,243.35) .. (344.74,249.7) -- (344.74,264.5) .. controls (344.74,270.85) and (349.89,276) .. (356.24,276) -- (356.24,276) .. controls (362.6,276) and (367.74,270.85) .. (367.74,264.5) -- cycle ;
		\draw   (348,243.17) -- (363.33,243.17) -- (363.33,253.17) -- (348,253.17) -- cycle ;
		\draw  [fill={rgb, 255:red, 62; green, 61; blue, 61 }  ,fill opacity=1 ] (348.07,256.7) .. controls (348.07,255.47) and (349.04,254.47) .. (350.24,254.47) .. controls (351.45,254.47) and (352.42,255.47) .. (352.42,256.7) .. controls (352.42,257.93) and (351.45,258.93) .. (350.24,258.93) .. controls (349.04,258.93) and (348.07,257.93) .. (348.07,256.7) -- cycle ;
		\draw  [fill={rgb, 255:red, 62; green, 61; blue, 61 }  ,fill opacity=1 ] (354.07,256.7) .. controls (354.07,255.47) and (355.04,254.47) .. (356.24,254.47) .. controls (357.45,254.47) and (358.42,255.47) .. (358.42,256.7) .. controls (358.42,257.93) and (357.45,258.93) .. (356.24,258.93) .. controls (355.04,258.93) and (354.07,257.93) .. (354.07,256.7) -- cycle ;
		\draw  [fill={rgb, 255:red, 62; green, 61; blue, 61 }  ,fill opacity=1 ] (360.07,256.7) .. controls (360.07,255.47) and (361.04,254.47) .. (362.24,254.47) .. controls (363.45,254.47) and (364.42,255.47) .. (364.42,256.7) .. controls (364.42,257.93) and (363.45,258.93) .. (362.24,258.93) .. controls (361.04,258.93) and (360.07,257.93) .. (360.07,256.7) -- cycle ;
		\draw  [fill={rgb, 255:red, 62; green, 61; blue, 61 }  ,fill opacity=1 ] (348.07,262.7) .. controls (348.07,261.47) and (349.04,260.47) .. (350.24,260.47) .. controls (351.45,260.47) and (352.42,261.47) .. (352.42,262.7) .. controls (352.42,263.93) and (351.45,264.93) .. (350.24,264.93) .. controls (349.04,264.93) and (348.07,263.93) .. (348.07,262.7) -- cycle ;
		\draw  [fill={rgb, 255:red, 62; green, 61; blue, 61 }  ,fill opacity=1 ] (354.07,262.7) .. controls (354.07,261.47) and (355.04,260.47) .. (356.24,260.47) .. controls (357.45,260.47) and (358.42,261.47) .. (358.42,262.7) .. controls (358.42,263.93) and (357.45,264.93) .. (356.24,264.93) .. controls (355.04,264.93) and (354.07,263.93) .. (354.07,262.7) -- cycle ;
		\draw  [fill={rgb, 255:red, 62; green, 61; blue, 61 }  ,fill opacity=1 ] (360.07,262.7) .. controls (360.07,261.47) and (361.04,260.47) .. (362.24,260.47) .. controls (363.45,260.47) and (364.42,261.47) .. (364.42,262.7) .. controls (364.42,263.93) and (363.45,264.93) .. (362.24,264.93) .. controls (361.04,264.93) and (360.07,263.93) .. (360.07,262.7) -- cycle ;
		\draw  [fill={rgb, 255:red, 62; green, 61; blue, 61 }  ,fill opacity=1 ] (348.07,268.7) .. controls (348.07,267.47) and (349.04,266.47) .. (350.24,266.47) .. controls (351.45,266.47) and (352.42,267.47) .. (352.42,268.7) .. controls (352.42,269.93) and (351.45,270.93) .. (350.24,270.93) .. controls (349.04,270.93) and (348.07,269.93) .. (348.07,268.7) -- cycle ;
		\draw  [fill={rgb, 255:red, 62; green, 61; blue, 61 }  ,fill opacity=1 ] (354.07,268.7) .. controls (354.07,267.47) and (355.04,266.47) .. (356.24,266.47) .. controls (357.45,266.47) and (358.42,267.47) .. (358.42,268.7) .. controls (358.42,269.93) and (357.45,270.93) .. (356.24,270.93) .. controls (355.04,270.93) and (354.07,269.93) .. (354.07,268.7) -- cycle ;
		\draw  [fill={rgb, 255:red, 62; green, 61; blue, 61 }  ,fill opacity=1 ] (360.07,268.7) .. controls (360.07,267.47) and (361.04,266.47) .. (362.24,266.47) .. controls (363.45,266.47) and (364.42,267.47) .. (364.42,268.7) .. controls (364.42,269.93) and (363.45,270.93) .. (362.24,270.93) .. controls (361.04,270.93) and (360.07,269.93) .. (360.07,268.7) -- cycle ;
		
		\draw (272.33,238.5) node  {\includegraphics[width=70.5pt,height=89.25pt]{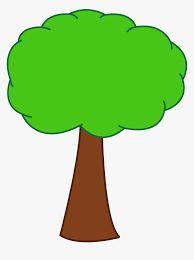}};
		\draw (387.17,153.5) node  {\includegraphics[width=47.75pt,height=101.25pt]{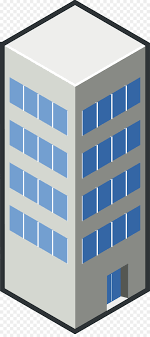}};
		\draw    (187.33,48) .. controls (186.99,50.33) and (185.65,51.33) .. (183.32,50.98) .. controls (180.99,50.64) and (179.65,51.64) .. (179.31,53.97) .. controls (178.97,56.3) and (177.63,57.3) .. (175.3,56.95) .. controls (172.97,56.6) and (171.63,57.6) .. (171.28,59.93) .. controls (170.94,62.26) and (169.6,63.26) .. (167.27,62.92) .. controls (164.94,62.57) and (163.6,63.57) .. (163.26,65.9) .. controls (162.92,68.23) and (161.58,69.23) .. (159.25,68.88) .. controls (156.92,68.54) and (155.58,69.54) .. (155.23,71.87) .. controls (154.89,74.2) and (153.55,75.2) .. (151.22,74.85) .. controls (148.89,74.51) and (147.55,75.51) .. (147.21,77.84) .. controls (146.87,80.17) and (145.53,81.17) .. (143.2,80.82) .. controls (140.87,80.47) and (139.53,81.47) .. (139.19,83.8) .. controls (138.84,86.13) and (137.5,87.13) .. (135.17,86.79) .. controls (132.84,86.44) and (131.5,87.44) .. (131.16,89.77) .. controls (130.82,92.1) and (129.48,93.1) .. (127.15,92.75) .. controls (124.82,92.41) and (123.48,93.41) .. (123.14,95.74) .. controls (122.79,98.07) and (121.45,99.07) .. (119.12,98.72) -- (117.36,100.03) -- (110.94,104.81) ;
		\draw [shift={(109.33,106)}, rotate = 323.37] [color={rgb, 255:red, 0; green, 0; blue, 0 }  ][line width=0.75]    (10.93,-3.29) .. controls (6.95,-1.4) and (3.31,-0.3) .. (0,0) .. controls (3.31,0.3) and (6.95,1.4) .. (10.93,3.29)   ;
		\draw  [dash pattern={on 0.75pt off 0.75pt}]  (472.33,94) .. controls (470.08,94.71) and (468.61,93.94) .. (467.9,91.69) .. controls (467.19,89.44) and (465.71,88.68) .. (463.46,89.39) .. controls (461.21,90.1) and (459.74,89.33) .. (459.03,87.08) .. controls (458.32,84.83) and (456.84,84.06) .. (454.59,84.77) .. controls (452.34,85.48) and (450.86,84.72) .. (450.15,82.47) .. controls (449.44,80.22) and (447.97,79.45) .. (445.72,80.16) .. controls (443.47,80.87) and (441.99,80.1) .. (441.28,77.85) .. controls (440.57,75.6) and (439.09,74.84) .. (436.84,75.55) .. controls (434.59,76.26) and (433.12,75.49) .. (432.41,73.24) .. controls (431.7,70.99) and (430.22,70.22) .. (427.97,70.93) .. controls (425.72,71.64) and (424.25,70.88) .. (423.54,68.63) .. controls (422.83,66.38) and (421.35,65.61) .. (419.1,66.32) .. controls (416.85,67.03) and (415.37,66.26) .. (414.66,64.01) .. controls (413.95,61.76) and (412.48,61) .. (410.23,61.71) .. controls (407.98,62.42) and (406.5,61.65) .. (405.79,59.4) .. controls (405.08,57.15) and (403.61,56.38) .. (401.36,57.09) .. controls (399.11,57.8) and (397.63,57.04) .. (396.92,54.79) .. controls (396.21,52.54) and (394.73,51.77) .. (392.48,52.48) .. controls (390.23,53.19) and (388.76,52.42) .. (388.05,50.17) .. controls (387.34,47.92) and (385.86,47.15) .. (383.61,47.86) -- (381.21,46.61) -- (374.11,42.92) ;
		\draw [shift={(372.33,42)}, rotate = 27.47] [color={rgb, 255:red, 0; green, 0; blue, 0 }  ][line width=0.75]    (10.93,-3.29) .. controls (6.95,-1.4) and (3.31,-0.3) .. (0,0) .. controls (3.31,0.3) and (6.95,1.4) .. (10.93,3.29)   ;
		\draw  [dash pattern={on 0.75pt off 0.75pt}]  (543.33,120) .. controls (541.05,120.6) and (539.61,119.76) .. (539.01,117.48) .. controls (538.42,115.2) and (536.98,114.36) .. (534.7,114.96) .. controls (532.42,115.56) and (530.98,114.72) .. (530.38,112.44) .. controls (529.78,110.16) and (528.34,109.32) .. (526.06,109.92) .. controls (523.78,110.52) and (522.34,109.68) .. (521.74,107.4) .. controls (521.14,105.12) and (519.7,104.28) .. (517.42,104.88) .. controls (515.14,105.48) and (513.7,104.64) .. (513.1,102.36) .. controls (512.5,100.08) and (511.06,99.24) .. (508.78,99.85) .. controls (506.5,100.45) and (505.06,99.61) .. (504.46,97.33) .. controls (503.86,95.05) and (502.42,94.21) .. (500.14,94.81) .. controls (497.86,95.41) and (496.42,94.57) .. (495.83,92.29) .. controls (495.23,90.01) and (493.79,89.17) .. (491.51,89.77) .. controls (489.23,90.37) and (487.79,89.53) .. (487.19,87.25) .. controls (486.59,84.97) and (485.15,84.13) .. (482.87,84.73) .. controls (480.59,85.33) and (479.15,84.49) .. (478.55,82.21) .. controls (477.95,79.93) and (476.51,79.09) .. (474.23,79.69) .. controls (471.95,80.29) and (470.51,79.45) .. (469.91,77.17) .. controls (469.31,74.89) and (467.87,74.05) .. (465.59,74.65) .. controls (463.31,75.25) and (461.87,74.41) .. (461.27,72.13) .. controls (460.68,69.85) and (459.24,69.01) .. (456.96,69.61) .. controls (454.68,70.21) and (453.24,69.37) .. (452.64,67.09) .. controls (452.04,64.81) and (450.6,63.97) .. (448.32,64.57) .. controls (446.04,65.17) and (444.6,64.33) .. (444,62.05) .. controls (443.4,59.77) and (441.96,58.93) .. (439.68,59.54) .. controls (437.4,60.14) and (435.96,59.3) .. (435.36,57.02) .. controls (434.76,54.74) and (433.32,53.9) .. (431.04,54.5) .. controls (428.76,55.1) and (427.32,54.26) .. (426.72,51.98) .. controls (426.12,49.7) and (424.68,48.86) .. (422.4,49.46) .. controls (420.12,50.06) and (418.68,49.22) .. (418.09,46.94) .. controls (417.49,44.66) and (416.05,43.82) .. (413.77,44.42) .. controls (411.49,45.02) and (410.05,44.18) .. (409.45,41.9) .. controls (408.85,39.62) and (407.41,38.78) .. (405.13,39.38) .. controls (402.85,39.98) and (401.41,39.14) .. (400.81,36.86) .. controls (400.21,34.58) and (398.77,33.74) .. (396.49,34.34) .. controls (394.21,34.94) and (392.77,34.1) .. (392.17,31.82) .. controls (391.57,29.54) and (390.13,28.7) .. (387.85,29.3) -- (383.97,27.04) -- (377.06,23.01) ;
		\draw [shift={(375.33,22)}, rotate = 30.26] [color={rgb, 255:red, 0; green, 0; blue, 0 }  ][line width=0.75]    (10.93,-3.29) .. controls (6.95,-1.4) and (3.31,-0.3) .. (0,0) .. controls (3.31,0.3) and (6.95,1.4) .. (10.93,3.29)   ;
		\draw  [dash pattern={on 0.75pt off 0.75pt}]  (350,211) .. controls (347.96,209.82) and (347.53,208.21) .. (348.72,206.17) .. controls (349.91,204.13) and (349.48,202.52) .. (347.44,201.33) .. controls (345.4,200.15) and (344.97,198.54) .. (346.16,196.5) .. controls (347.35,194.46) and (346.92,192.85) .. (344.88,191.67) .. controls (342.84,190.48) and (342.41,188.87) .. (343.6,186.83) .. controls (344.79,184.79) and (344.36,183.18) .. (342.32,182) .. controls (340.28,180.82) and (339.85,179.21) .. (341.04,177.17) .. controls (342.23,175.13) and (341.8,173.52) .. (339.76,172.33) .. controls (337.72,171.15) and (337.29,169.54) .. (338.48,167.5) .. controls (339.67,165.46) and (339.24,163.85) .. (337.2,162.67) .. controls (335.16,161.48) and (334.73,159.87) .. (335.92,157.83) .. controls (337.11,155.79) and (336.68,154.18) .. (334.64,153) .. controls (332.6,151.82) and (332.17,150.21) .. (333.36,148.17) .. controls (334.55,146.13) and (334.12,144.52) .. (332.08,143.33) .. controls (330.04,142.15) and (329.61,140.54) .. (330.8,138.5) .. controls (331.99,136.46) and (331.56,134.85) .. (329.52,133.66) .. controls (327.48,132.48) and (327.05,130.87) .. (328.24,128.83) .. controls (329.43,126.8) and (329,125.19) .. (326.97,124) .. controls (324.93,122.81) and (324.5,121.2) .. (325.69,119.16) .. controls (326.88,117.12) and (326.45,115.51) .. (324.41,114.33) .. controls (322.37,113.15) and (321.94,111.54) .. (323.13,109.5) .. controls (324.32,107.46) and (323.89,105.85) .. (321.85,104.66) .. controls (319.81,103.48) and (319.38,101.87) .. (320.57,99.83) .. controls (321.76,97.79) and (321.33,96.18) .. (319.29,95) .. controls (317.25,93.81) and (316.82,92.2) .. (318.01,90.16) .. controls (319.2,88.12) and (318.77,86.51) .. (316.73,85.33) .. controls (314.69,84.15) and (314.26,82.54) .. (315.45,80.5) .. controls (316.64,78.46) and (316.21,76.85) .. (314.17,75.66) .. controls (312.13,74.48) and (311.7,72.87) .. (312.89,70.83) -- (312.85,70.67) -- (310.8,62.93) ;
		\draw [shift={(310.29,61)}, rotate = 75.17] [color={rgb, 255:red, 0; green, 0; blue, 0 }  ][line width=0.75]    (10.93,-3.29) .. controls (6.95,-1.4) and (3.31,-0.3) .. (0,0) .. controls (3.31,0.3) and (6.95,1.4) .. (10.93,3.29)   ;
		\draw  [fill={rgb, 255:red, 0; green, 0; blue, 0 }  ,fill opacity=1 ] (488.33,142.67) .. controls (488.33,141.38) and (489.38,140.33) .. (490.67,140.33) .. controls (491.96,140.33) and (493,141.38) .. (493,142.67) .. controls (493,143.96) and (491.96,145) .. (490.67,145) .. controls (489.38,145) and (488.33,143.96) .. (488.33,142.67) -- cycle ;
		\draw  [fill={rgb, 255:red, 0; green, 0; blue, 0 }  ,fill opacity=1 ] (508.33,142.67) .. controls (508.33,141.38) and (509.38,140.33) .. (510.67,140.33) .. controls (511.96,140.33) and (513,141.38) .. (513,142.67) .. controls (513,143.96) and (511.96,145) .. (510.67,145) .. controls (509.38,145) and (508.33,143.96) .. (508.33,142.67) -- cycle ;
		\draw  [fill={rgb, 255:red, 0; green, 0; blue, 0 }  ,fill opacity=1 ] (525.33,143.67) .. controls (525.33,142.38) and (526.38,141.33) .. (527.67,141.33) .. controls (528.96,141.33) and (530,142.38) .. (530,143.67) .. controls (530,144.96) and (528.96,146) .. (527.67,146) .. controls (526.38,146) and (525.33,144.96) .. (525.33,143.67) -- cycle ;
		
		\draw (135,130) node [anchor=north west][inner sep=0.75pt]   [align=left] {M};
		\draw (63,131) node [anchor=north west][inner sep=0.75pt]   [align=left] {2};
		\draw (34,131) node [anchor=north west][inner sep=0.75pt]   [align=left] {1};
		\draw (114,48) node [anchor=north west][inner sep=0.75pt]   [align=left] {\textbf{H}};
		\draw (124,54) node [anchor=north west][inner sep=0.75pt]   [align=left] {{\footnotesize 1}};
		\draw (280,112) node [anchor=north west][inner sep=0.75pt]   [align=left] {\textbf{h}};
		\draw (289,120) node [anchor=north west][inner sep=0.75pt]  [font=\footnotesize] [align=left] {2,1};
		\draw (381,55) node [anchor=north west][inner sep=0.75pt]   [align=left] {\textbf{h}};
		\draw (390,63) node [anchor=north west][inner sep=0.75pt]  [font=\footnotesize] [align=left] {2,2};
		\draw (457,55) node [anchor=north west][inner sep=0.75pt]   [align=left] {\textbf{h}};
		\draw (466,63) node [anchor=north west][inner sep=0.75pt]  [font=\footnotesize] [align=left] {2,K};
		\draw (209,4) node [anchor=north west][inner sep=0.75pt]   [align=left] {{\small 1}};
		\draw (234,5) node [anchor=north west][inner sep=0.75pt]  [font=\small] [align=left] {2};
		\draw (352,53) node [anchor=north west][inner sep=0.75pt]  [font=\small] [align=left] {N};
	\end{tikzpicture}}
	\caption{{System model}} \label{sys_model}
\end{figure*}

We consider the uplink communication of a multi-user SIMO system where $K$ single antenna users communicate to an $M$-antenna BS with the help of an $N$ element RIS array. Due to natural or man-made obstacles, no direct link will be available between the $K$ users and the BS.  In such a scenario, we utilize a strategically placed RIS to manipulate the phase of the incoming signal to improve the SINR received by the BS. The signal received along the RIS is expected to have a larger SINR compared to the other scattered components \cite{bjornson2022reconfigurable}. Thus, we do not consider any link between the BS and the users in this analysis. Such a model is widely adopted for the study of RIS-aided communication systems \cite{zhang2021reconfigurable,zeng2021reconfigurable,qiao2020secure,huang2019reconfigurable}. The signal received at the BS is given by 
\begin{equation}
\vec{y} = \left( \vec{H}_1 \vec{R}_{\text{RIS}}^{\frac{1}{2}} \vec{\Phi} \mathbf{R}_{\text{users}}\vec{H}_2  \right) \vec{x} + \vec{n},
\end{equation}
where $\vec{H}_1 \in \mathbb{C}^{M \times N}$ represents the LOS channel matrix between the $\mathrm{BS}$ and {$\mathrm{RIS}, \mathbf{R}_{\text{RIS}} \in \mathbb{C}^{N \times N}$, $\vec{R}_{\text{users}} \in \mathbb{C}^{N \times N}$ 
represents the spatial correlation matrix of the RIS elements} and the users respectively, $\vec{H}_2 \in \mathbb
{C}^{N \times K}$ represents the channel between the $K$ users and the RIS, i.e $\mathbf{{h}}_{2, k}=\ell_k \mathbf{\tilde{h}}_{2, k}$ where $\ell_k$ is the path loss corresponding to the link between user $k$ and the RIS and $ \mathbf{\tilde{h}}_{2, k} \sim \mathcal{C} \mathcal{N}\left(\mathbf{0}, \mathbf{I}_{N}\right) \in \mathbb{C}^{N \times 1}$ denotes the fading channel gain between the RIS and user $k$, $\Vec{x}$ represents the transmit signal vector and $n_{k}$ represents the thermal noise, modeled as a $\mathcal{C} \mathcal{N}\left(0, \tilde{\sigma}^{2}\right)$ random variable $(\mathrm{RV}) .$ Furthermore, we have $\vec{\Phi}=\operatorname{diag}\left( \phi_{1}, \ldots, \phi_{N}\right) \in$ $\mathbb{C}^{N \times N}$ is the diagonal matrix accounting for the response of the RIS elements, and $\phi_{n}=\exp \left(j \theta_{n}\right), n\in \{1, \ldots, N\}$, where $\theta_n$ is the phase shift induced by the $n$-th RIS element. Note that the received signal can be rewritten as 
\begin{equation}
\vec{y} = \sum\limits_{k=1}^K p_k \vec{H}_1 \vec{R}_{\text{RIS}}^{\frac{1}{2}} \vec{\Phi} \mathbf{R}_{\text{users}} \vec{h}_{2,k} x_k + \vec{n}, 
\end{equation}
where $x_k$ represents the complex symbol transmitted from user $k$ with transmit power $p_k$. Now, let $\boldsymbol {\beta} \in \mathbb{C}^{K \times M }$ be the BF matrix at the BS, and the received symbols can be decoded as 
\begin{equation}
    \vec{r} = \boldsymbol {\beta} \vec{y}.
\end{equation}
Then, the decoded symbol from user $k$ can be represented as 
\begin{equation}
    r_k = \boldsymbol {\beta}_k^H \vec{y},
\end{equation}
where $\boldsymbol {\beta}_k$ is the $k$-th row of the matrix $\boldsymbol{\beta}$. The SINR for user $k$ at the BS is hence given by 
\begin{align}
& \text{SINR}_k = \nonumber \\ &  \frac{\frac{p_k}{K}\left|\boldsymbol {\beta}_k^H\vec{H}_1 \vec{R}_{\text{RIS}}^{\frac{1}{2}} \vec{\Phi} \mathbf{R}_{\text{users}} \vec{h}_{2,k} \right|^2}{\sum\limits_{i=1,i \neq k}^K \frac{p_i}{K} \left|\boldsymbol {\beta}_k^H\vec{H}_1 \vec{R}_{\text{RIS}}^{\frac{1}{2}} \vec{\Phi} \mathbf{R}_{\text{users}} \vec{h}_{2,i} \right|^2 + {\sigma^2 ||\boldsymbol {\beta}_k||^2}},
\end{align}
where $\sigma^2=\frac{\tilde{\sigma^2}}{K}$. The SINR experienced by each user is thus a function of the power allocation, BF at the BS, and the phase shift induced by the RIS elements. To ensure a fair service for all users, we would like to jointly optimize these variables such that the minimum SINR in the system is maximized, i.e., maximizing $\min\limits_{k \in \mathcal{K}} \text{SINR}_k$. In the next section, we formulate an optimization problem to identify the optimal power allocation vector, BF vectors at the BS and phase shifts introduced by the RIS elements such that the constraints on the transmit power of each user are satisfied. We also discuss how the studied optimization problem can be used for the resource allocation of RIS-aided communication systems with constraints on the maximum EMF of each user.  

\section{Optimization Problem and solution} \label{opti_prob}

The optimization problem discussed in the last section can be mathematically formulated as follows:
\begin{equation}
\tag{$\mathcal{P}_1$}
\begin{aligned}
\max_{\{p_k\},\vec{\Phi}, \boldsymbol{\beta}} \quad & \min\limits_k \ \text{SINR}_k \\
\textrm{s.t.} \quad &0 \leq  \frac{p_k}{K} \leq p_{\text{max},k},  \ \forall  k \in \mathcal{K} \label{prob1}\\
  & ||\boldsymbol{\beta}_k||=1,  \ \forall  k \in \mathcal{K},\\
  &|\phi_n| =1, \ \forall  n \in \{1,\cdots,N\}.
\end{aligned}
\end{equation}
where $\mathcal{K}=\{1,2,\cdots,K\}$.
Note that we have also imposed constraints on the norm of each BF vector and the magnitudes of the phase shift at each of the RIS elements. The solution for the above optimization problem depends upon the instantaneous channel gains, and hence the BF, power allocation vector, and phase shift vector at the RIS need to be updated for every different channel realization, i.e., after the channel coherence time. Hence, the optimal RIS phase shift needs to be evaluated at the BS for every channel realization, and there needs to be a feedback scheme to send this information to the RIS. Such frequent feedbacks are costly, and hence in this work, we propose to design the phase shift vector at the RIS using statistical CSI. {Note that the joint optimization of the BF matrix, the power variables and the RIS phase shift vector is not a convex problem and hence does not admit a very simple solution. We propose using the simple alternating optimization technique wherein the joint optimization problem is broken down into multiple sub-problems, which are solved independently. We shall then iterate over the solutions for each problem until the solution converges. For BF and power allocation, we follow the same approach as \cite{emfrisinst}, and the corresponding solutions are discussed in the next paragraph. }
\subsection{BF and power allocation}\label{bf_power_alloc}
The optimal BF vector is given by the solution for the following problem.
\begin{equation}
\tag{$\mathcal{SP}_1$}
\begin{aligned}
\max_{\boldsymbol{\beta}} \quad & \min\limits_k \ \text{SINR}_k \\
\textrm{s.t.} \ &  ||\boldsymbol{\beta}_k||=1,  \ \forall  k \in \mathcal{K}.
\end{aligned}
\end{equation}
Following the steps in \cite{emfrisinst}, the solution for the above problem is given by
\begin{equation}
\boldsymbol{\beta}_k = \frac{\left( \boldsymbol{\Sigma}_k + \sigma^2 \mathbf{I}_{M} \right)^{-1} \vec{g}_k}{\left\vert\left\vert{\left( \boldsymbol{\Sigma}_k + \sigma^2 \mathbf{I}_{M} \right)^{-1} \vec{g}_k}\right \vert\right \vert},
\label{beta_opti}
\end{equation}
where $\boldsymbol{\Sigma}_k=\sum\limits_{i=1,i\neq k}^K \frac{p_i}{K} \vec{g}_i \vec{g}_i^H $ and $\vec{g}_k = \vec{H}_1 \vec{R}_{\text{RIS}}^{\frac{1}{2}} \vec{\Phi}\mathbf{R}_{\text{users}} \vec{h}_{2,k}$. For this choice of $\{\boldsymbol{\beta}_k\}$, the SINR of the $k$-th user is given by
\begin{equation}
\text{SINR}_k = \frac{p_k}{K} \vec{g}_k^H \left( \boldsymbol{\Sigma}_k + \sigma^2 \mathbf{I}_{M} \right)^{-1} \vec{g}_k.
\end{equation}
Similarly, the sub-problem of identifying the optimal transmit power policy can be formulated as 
\begin{equation}
\begin{aligned}
\max_{\{p_k\}} \quad & \min\limits_k \ \text{SINR}_k \\
\textrm{s.t.} \quad &0 \leq  \frac{p_k}{K} \leq p_{\text{max},k},  \ \forall  k \in \mathcal{K}.
\end{aligned}
\end{equation}
Following steps similar to \cite[Eq.~(2)]{cai2011unified}, this problem can be rewritten as 

\begin{align}\tag{$\mathcal{SP}_2$}
\min_{ \{p_k\},\tau} &  & \hspace{-20mm} \tau^{-1} & & \\
\text { s.t. } & & \frac{\tau \left(\sum\limits_{i=1,i \neq k}^K \frac{p_i}{K} f_{k,i} + n_k \right)}{p_k f_{k,k}}  \leq 1, \nonumber \ \ &  \forall  k \in \mathcal{K},\\
  & & 0 \leq  \frac{p_k}{K} \leq p_{\text{max},k},    \ \ &    \forall  k \in \mathcal{K}, \nonumber 
\end{align}
where $f_{k,i}=\left|\boldsymbol{\beta}_k^H\vec{H}_1 \vec{R}_{\text{RIS}}^{\frac{1}{2}} \vec{\Phi}\mathbf{R}_{\text{users}} \vec{h}_{2,i} \right|^2$ and $n_k=\sigma^2$. The above problem can be solved using the popular technique of geometric programming. Geometric programming solvers are available in popular optimization packages like CVX and can be solved efficiently \cite{cvx}.

\subsection{Phase shift design using asymptotic statistics}

In this section, we study phase shift design using the deterministic equivalent of the minimum SINR instead of the exact expression for the minimum. We study the following two approaches here:

\begin{itemize}
\item[1.] \textit{Approach 1}: BS has full CSI and performs the BF optimization, and both power allocation and RIS phase shift design are performed using statistical CSI.
\item[2.] \textit{Approach 2}: BF and power allocation are performed using exact CSI, and only phase shift design is performed using statistical CSI.
\end{itemize}
To set the stage for both designs, we will carry out an asymptotic analysis of the users' SINR and their allocated powers.
Now, using the optimal BF vectors in \eqref{beta_opti}, the power allocation is achieved by solving the optimization problem in $\mathcal{P}_2$. In \cite{tan2014wireless}, the authors addressed a similar problem to study max-min fairness optimization in wireless networks. Using Lemma 1 from \cite{tan2014wireless}, we observe that the optimal solution for $\mathcal{P}_2$ would ensure that all the users experience the same SINR, say $\tau^*$. Thus, we have,
\begin{equation}
\tau^* = \text{SINR}_k = \frac{{p}_k}{K} \vec{g}_k^H \left( {\Sigma}_k + {\sigma}^2 \mathbf{I}_{M} \right)^{-1} \!\vec{g}_k,  \ \forall  k \in \mathcal{K}.
\label{opti_sinr_power}
\end{equation}

{Let us rewrite $\tau^*$ as $\tau^* = \frac{{p}_k}{K}\ell_k \tilde{\vec{g}}_k^H \left( {\Sigma}_k + \sigma^2 \mathbf{I}_{M} \right)^{-1} \vec{\tilde{g}}_k $, where $\tilde{\vec{g}}_k=\underbrace{\vec{H}_1\vec{R}_{\text{RIS}}^{\frac{1}{2}} \vec{\Phi} 
\vec{R}_{\text{users}}^{\frac{1}{2}}}_{\vec{U}} \tilde{\vec{h}}_{2,k}$, and $\vec{g}_k = \sqrt{\ell_k} \tilde{\vec{g}}_k$. }
Letting $d_k:= \frac{1}{K} \tilde{\vec{g}}_k^H \left( {\Sigma}_k + \sigma^2 \mathbf{I}_{M} \right)^{-1} \vec{\tilde{g}}_k=\frac{1}{K} \tilde{\vec{h}}_{2,k}^H \vec{U}^H \left( {\Sigma}_k + \sigma^2 \mathbf{I}_{M} \right)^{-1} \vec{U} \tilde{\vec{h}}_{2,k}$, we obtain

\begin{align}
\tau^* =  
p_k\ell_k d_k \label{eq:tau}
\end{align}
{or equivalently $p_k=\frac{\tau^*}{\ell_kd_k}$. }

{Using \eqref{eq:tau}, we can see that $d_1,\cdots,d_K$ are  solutions to the following system of equations:
\begin{equation}
{ d}_k= h_k(d_1,\cdots,d_K;\tau^\star),  \ \forall  k \in \mathcal{K}, \label{eq:r}
\end{equation}
where $h_k(d_1,\cdots,d_K;\tau):\mathbb{R}^{K}\times \mathbb{R}:  (d_1,\cdots,d_K)\mapsto \frac{1}{K}\tilde{\bf g}_k^{H}\left(\sum_{i\neq k} \frac{\tau }{ d_i}\tilde{\bf g}_i\tilde{\bf g}_i^{H}+\sigma^2{\bf I}\right)^{-1}\tilde{\bf g}_k$.
To continue, we exploit the fact that $${h}:\mathbb{R}^{K} \to {\mathbb{R}}^{K}: (d_1,\cdots,d_K)\mapsto h_k(d_1,\cdots,d_K;\tau)$$ is a standard interference function. Hence, the existence and uniqueness of the solution to \eqref{eq:r} hold true for any $\tau\geq 0$. Particularly, letting $\tau\geq 0$, the following system of equations:
\begin{equation}
{ d}_k= h_k(d_1,\cdots,d_K;\tau),  \ \forall  k \in \mathcal{K}, \label{eq:dktau}
\end{equation}
admits a unique solution which we denote by $d_1(\tau),\cdots,d_K(\tau)$. With this, we may prove that $\mathcal{P}_2$ amounts to solving the following problem.
\begin{align}\tag{$\mathcal{P}_{\rm eq}$}
	\min_{d_k,\tau} &  \ \  \frac{1}{\tau}  \\
	\text { s.t. } &  \ \ d_k(\tau)=h_k(d_1,\cdots,d_K;\tau), \ \  \forall  k \in \mathcal{K},\\
	 & \ \ \frac{\tau}{d_k(\tau)}\leq K\,\ell_k\, p_{{\rm max},k},  \ \  \forall  k \in \mathcal{K}. 
\end{align}
The equivalent problem $\mathcal{P}_{\rm eq}$ is not convex. However,  when it comes to asymptotic analysis, it is much simpler to handle. Prior to stating our main results, we shall first introduce the growth rate regime considered in our asymptotic analysis. Particularly, we will assume the following assumptions. 

\textit{Assumption 1:} $ M, N$ and $K$ grow large with a bounded ratio as $0<\lim \inf \frac{K}{M} \leq \lim \sup \frac{K}{M}<1$ and $0<\lim \inf \frac{N}{M} \leq$ $\lim \sup \frac{N}{M}<\infty$.
In the sequel, this assumption is denoted as $\underset{M \rightarrow \infty}{\longrightarrow}$.

\textit{Assumption 2:} The sequence of correlation matrices $({\bf R}_{\text{RIS}})_{N}$, $({\bf R}_{\text{users}})_{N}$ and the channel matrices $({\bf H}_1)_M$ satisfy:
$$
\limsup_{M,N,K} \max\left(\|{\bf R}_{\text{RIS}}\|, \|{\bf R}_{\text{users}}\|,\|{\bf H}_1\|\right)<\infty.
$$
Moreover, the sequence of maximum powers $p_{{\rm max},k}$ and link path losses $\ell_k$, $k=1,\cdots,K$ satisfy
$$
\limsup_{M,N,K} \max_{1\leq k \leq K} K \ell_k p_{{\rm max},k} <\infty, 
$$
and
$$
\liminf_{M,N,K} \max_{1\leq k \leq K} K \ell_k p_{{\rm max},k} >0. 
$$

\begin{theorem}
	Consider the regime in Assumption 1 and Assumption 2. For $\tau\geq 0$, let $d_1(\tau),\cdots,d_K(\tau)$ be the unique solutions to \eqref{eq:dktau}. Then, the following equation:
	$$
	d=\frac{1}{K}\mathrm{Trace} \left \lbrace  \vec{U}\vec{U}^H \left(\vec{U}\vec{U}^H  \frac{\tau}{{d}(1+\tau)}+ {\sigma}^2 \mathbf{I}_{M} \right)^{-1} \right \rbrace
	$$
	admits a unique solution $\overline{d}(\tau)$.  Moreover, for any compact interval $\mathcal{I}\subset(0,\infty)$, we have:
	$$
	0<\inf_{\tau\in\mathcal{I}}\overline{d}(\tau)\leq\sup_{\tau\in\mathcal{I}}\overline{d}(\tau)<\infty
	$$
	and
	$$
	\sup_{\tau\in\mathcal{I}}\max_{1\leq k \leq K} \left|d_k(\tau)-\overline{d}(\tau)\right| \underset{M \rightarrow \infty}{\longrightarrow} 0.
	$$
	\label{th:1}
\end{theorem}
\begin{proof}
	The proof of Theorem \ref{th:1} is deferred to Appendix \ref{app:A}.
\end{proof}
Theorem \ref{th:1} studies the asymptotic behaviour of fixed point equations of standard interference functions in the form of \eqref{eq:dktau}. We believe that it can be used beyond the context of our work to handle, for instance, constraints expressed as linear combinations of user powers. For this reason, we formulate it as a stand-alone result to facilitate its use in more general settings. With Theorem \ref{th:1} at hand, we are now ready to state our results regarding the asymptotic convergence of the per-user SINR and the per-user powers. 
\begin{theorem}
Consider the setting of Theorem \ref{th:1}, let $\alpha_{0}:=\min_{1\leq k\leq K} \left\{K\ell_kp_{{\rm max},k}\right\}$, then, the following equation:
\begin{equation}
    \tau=\frac{\alpha_{0}}{K}\mathrm{Trace} \left \lbrace  \vec{U}\vec{U}^H \left({\vec{U}\vec{U}^H}  \frac{\alpha_{0}}{(1+\tau)}+ {\sigma}^2 \mathbf{I}_{M} \right)^{-1} \right \rbrace,
    \label{eq:tau_bar}
\end{equation}
	admits a unique solution $\overline{\tau}$. Let $\overline{d}$ be the unique solution to the following equation:
		$$
	d=\frac{1}{K}\mathrm{Trace} \left \lbrace  \vec{U}\vec{U}^H \left(\vec{U}\vec{U}^H  \frac{\overline{\tau}}{{d}(1+\overline{\tau})}+ {\sigma}^2 \mathbf{I}_{M} \right)^{-1} \right \rbrace.
	$$
	Then, the following convergences hold true:
	\begin{align}
	&\tau^\star-\overline{\tau}\underset{M \rightarrow \infty}{\longrightarrow} 0, \\
	& \max_{1\leq k \leq K} \left|d_k-\overline{d}\right| \underset{M \rightarrow \infty}{\longrightarrow} 0,\\ 
	& \max_{1\leq k \leq K} \left|p_k-\frac{\alpha_0}{\ell_k} \right| \underset{M \rightarrow \infty}{\longrightarrow} 0.
	\end{align}
where $d_1,\cdots,d_K$ are the unique solutions to Problem $(\mathcal{P}_{\rm eq})$.  \label{th:th_2}
\end{theorem}
\begin{proof}
	See Appendix \ref{app:B}.
\end{proof}
}

\subsubsection{RIS Phase optimization using the deterministic equivalents} \label{asymp_phase_opti}
Given that we have derived the asymptotic deterministic equivalent of $\tau^*$, we now use this to derive the optimal values of RIS phase shifts that maximize the minimum SINR. Thus, the new optimization problem can be written as follows:

\begin{align}\label{opti_phase_stat_csi}
\tag{$\mathcal{P}_3$}
\max _{\Phi} &  \quad  \bar{\tau}=\frac{\alpha_{0}}{K} \mathrm{Trace} \left \lbrace  \vec{U}\vec{U}^H \left(\frac{\alpha_{0}}{(1+\bar{\tau})}\vec{U}\vec{U}^H  + \tilde{\sigma}^2 \mathbf{I}_{M} \right)^{-1} \right \rbrace \\
 \text{s.t} & \quad |\phi_n|=1, \ \forall n=\{1,\cdots,N\}. \nonumber 
\end{align}
We use the projected gradient descent to solve the above optimization problem. We begin by evaluating the derivative of $\bar{\tau}$ with respect to $\phi_n$ for all $n \in \{1,\cdots,N\}$ using the implicit function theorem. 
Let
\begin{equation}
g(\bar{\tau},\phi) = \bar{\tau}-\frac{\alpha_{0}}{K} \mathrm{Trace} \left \lbrace  \vec{U}\vec{U}^H \vec{T} \right \rbrace ,
\end{equation}
where $\vec{T}=\left(\frac{\alpha_{0}}{(1+\bar{\tau})}\vec{U}\vec{U}^H  + \tilde{\sigma}^2 \mathbf{I}_{M} \right)^{-1}$, thus, we have
\begin{equation}
\frac{\partial \bar{\tau}}{\partial \phi_{n}}=-\frac{\frac{\partial g}{\partial \phi_{n}}}{\frac{\partial g}{\partial \bar{\tau}}},
\end{equation}
where
\begin{equation}
\frac{\partial g}{\partial \bar{\tau}}=1-\frac{\alpha_{0}^2}{K(1+\bar{\tau})^{2}} \operatorname{Trace} \left \lbrace  \vec{U}\vec{U}^H \mathbf{T}\vec{U}\vec{U}^H \mathbf{T} \right \rbrace .
\end{equation}

Evaluating $\frac{\partial g}{\partial \phi_{n}}$, we have
\begin{align}
\frac{\partial g}{\partial \phi_n}& =\frac{\alpha_{0}}{K} \left \lbrace  \frac{\alpha_{0}}{1+\bar{\tau}} \left[\vec{R}_{\text{RIS}}^{\frac{1}{2}} \Vec{H}_1^H\vec{T}\Vec{R} \vec{T}\Vec{H}_1 \vec{R}_{\text{RIS}}^{\frac{1}{2}} \Vec{\Phi} \vec{R}_{\text{users}}  \right]_{n,n} - \right. \nonumber  \\ & \left.  \left[\vec{R}_{\text{RIS}}^{\frac{1}{2}} \Vec{H}_1^H\vec{T}\Vec{H}_1 \vec{R}_{\text{RIS}}^{\frac{1}{2}} \Vec{\Phi} \vec{R}_{\text{users}}  \right]_{n,n}  \right \rbrace. 
\label{gradient1}
\end{align}
Using (\ref{gradient1}), we implement the projected gradient descent to solve $\mathcal{P}_3$. Note that the solution does not depend on the instantaneous channel gains; hence, the phase optimization must only be repeated when the large-scale channel statistics vary. 

\subsubsection{Power allocation using asymptotic statistics} \label{asymp_power_opti}
\begin{table*}[t!]
	\centering
	\caption{CSI assumptions for BF, power allocation, and phase shift designs in different schemes compared}
	\label{simu_schemes}
	\begin{tabular}{|l|l|l|l|}
		\hline
		& \begin{tabular}[c]{@{}l@{}}BF\\ \end{tabular}                                                        & \begin{tabular}[c]{@{}l@{}}Power\\ Allocation\end{tabular}                                                    & \begin{tabular}[c]{@{}l@{}}RIS\\ Phases\end{tabular}                                                          \\ \hline
		Scheme 1 & \begin{tabular}[c]{@{}l@{}}Instantaneous\\ (sub-section \ref{bf_power_alloc})\end{tabular} & \begin{tabular}[c]{@{}l@{}}Instantaneous\\ (sub-section \ref{bf_power_alloc})\end{tabular} & \begin{tabular}[c]{@{}l@{}}Instantaneous\\ \cite{emfrisinst}\end{tabular}                    \\ \hline
		Scheme 2 & \begin{tabular}[c]{@{}l@{}}Instantaneous\\ (sub-section \ref{bf_power_alloc})\end{tabular} & \begin{tabular}[c]{@{}l@{}}Instantaneous\\ (sub-section \ref{bf_power_alloc})\end{tabular} & \begin{tabular}[c]{@{}l@{}}Statistical\\ (subsection \ref{asymp_phase_opti})\end{tabular}  \\ \hline
		Scheme 3 & \begin{tabular}[c]{@{}l@{}}Instantaneous\\ (sub-section \ref{bf_power_alloc})\end{tabular} & \begin{tabular}[c]{@{}l@{}}Statistical\\ (sub-section \ref{asymp_power_opti})\end{tabular} & \begin{tabular}[c]{@{}l@{}}Statistical\\ (subsection \ref{asymp_phase_opti})\end{tabular}  \\ \hline
		Scheme 4 & \begin{tabular}[c]{@{}l@{}}Instantaneous\\ (sub-section \ref{bf_power_alloc})\end{tabular} & \begin{tabular}[c]{@{}l@{}}Statistical\\ (sub-section \ref{asymp_power_opti})\end{tabular} & \begin{tabular}[c]{@{}l@{}}$\theta_n=0$\\ $\forall \ n \in \{1,\cdots,N\}$\end{tabular}                       \\ \hline
		Scheme 5 & \begin{tabular}[c]{@{}l@{}}Instantaneous\\ (sub-section \ref{bf_power_alloc})\end{tabular} & \begin{tabular}[c]{@{}l@{}}Instantaneous\\ (sub-section \ref{bf_power_alloc})\end{tabular} & \begin{tabular}[c]{@{}l@{}}$\theta_n=0$\\ $\forall \ n \in \{1,\cdots,N\}$\end{tabular}                       \\ \hline
		Scheme 6 & \begin{tabular}[c]{@{}l@{}}Instantaneous\\ (sub-section \ref{bf_power_alloc})\end{tabular} & \begin{tabular}[c]{@{}l@{}}Instantaneous\\ (sub-section \ref{bf_power_alloc})\end{tabular} & \begin{tabular}[c]{@{}l@{}}$\theta_n \sim \text{Unif}[0,2\pi]$\\ $\forall \ n \in \{1,\cdots,N\}$\end{tabular} \\ \hline
	\end{tabular}

\end{table*}
Note that along with the derivation for the deterministic equivalent for $\tau^*$, we have also established that $\frac{\bar{\tau}}{\bar{d}}= {p}_k \ell_k = \alpha_{0}$. From this relation, we can observe that asymptotically, the power allocation for the $k$-th user converges to
\begin{equation}
\overline{p}_k = \frac{\alpha_{0}}{ \ell_k},
\label{power_asymp}
\end{equation}
where we recall that $\alpha_{0} = \min \left \lbrace \ell_k p_{\text{max},k} \right \rbrace $.
In the next section, we also compare the performance of this power allocation scheme with the scheme discussed in subsection \ref{bf_power_alloc}.

\section{{EMF aware design use case}} \label{application}

This section presents one interesting application where the proposed resource allocation policy can be utilized. In problem $\mathcal{P}_1$, we would like to include the additional constraint that the EMF of each user is below a permissible limit. The EMF of the $k$-th user transmitting with a power ${p}_k$ can be evaluated as $\text{SAR}_{\text{ref},k}{p}_k$ where $\text{SAR}_{\text{ref},k}$ is the SAR for user $k$ when the transmitted power is unity. The max-min SINR optimization with constraints on the user's exposure can be expressed as follows:
\begin{equation}
\tag{$\mathcal{P}_2$}
\begin{aligned}
\max_{\{{p}_k\},\vec{\phi}, \vec{\beta}} \quad & \min\limits_k \ \text{SINR}_k \label{prob1}\\
\textrm{s.t.} \quad &0 \leq  \text{SAR}_{\text{ref},k} p_k \leq \text{SAR}_{\text{max},k} ,  \ \forall  k \in \mathcal{K},  \\
  &0 \leq p_k \leq \tilde{p}_{\text{max}},  \ \forall  k \in \mathcal{K}, \\
  & ||\boldsymbol{\beta}_k||=1,  \ \forall  k \in \mathcal{K},\\
  &|\phi_n| =1,  \ \forall n \in \{1,\cdots,N\}.
\end{aligned}
\end{equation}
Here, $p_{\text{max}}$ is the maximum transmit power of each user, and $\text{SAR}_{\text{max},k}$ is the maximum allowed exposure for user $k$. Note that the two constraints on the power variable can be combined and expressed as $0\leq p_k \leq \min \left \lbrace \tilde{p}_{\text{max}},\frac{\text{SAR}_{\text{max},k}}{\text{SAR}_{\text{ref},k}}\right \rbrace $. Let $\min \left \lbrace \tilde{p}_{\text{max}},\frac{\text{SAR}_{\text{max},k}}{\text{SAR}_{\text{ref},k}}\right \rbrace:=\tilde{p}_{\text{max},k}$. Then, we can notice that the problem in $\mathcal{P}_2$ is equivalent to the one in $\mathcal{P}_1$ for $\tilde{p}_{\text{max},k}=Kp_{\text{max},k}$, and hence can be solved using the method proposed in the previous section. In the next section, we present simulation results for this application and hence demonstrate the performance of the proposed solution methodology.   

\section{Simulation Results} \label{simu}

In this section, we consider a simulation setup similar to that used by the authors of \cite{kammoun2020asymptotic}. The coordinates of the BS and the RIS are given by $(0,0,10)$ and $(10,10,15)$, respectively. The coordinates of the user locations are given by $(x_k,y_k,1.5)$ where $x_k \sim \text{unif}\left[ 10, 15\right]$ and $y_k \sim \text{unif}\left[ 5, 10\right]$. One sample realization of the BS, RIS and the users for $K=10$ is shown in Fig. \ref{sim_fig_1}.  

\begin{figure}[t!]
    \centering
    \definecolor{mycolor1}{rgb}{0.00000,0.44700,0.74100}
\definecolor{mycolor2}{rgb}{0.85000,0.32500,0.09800}
\definecolor{mycolor3}{rgb}{0.92900,0.69400,0.12500}

	\begin{tikzpicture}
		\begin{axis}[%
			width=0.95\linewidth, 
			height=0.75\linewidth, 
			xmin=0,
			xmax=70,
			xtick={0, 10, 20, 30, 40, 50},
			tick align=outside,
			xlabel={$x$},
			ymin=0,
			ymax=60,
			ytick={0, 20,  40, 60},
			ylabel={$y$},
			zmin=0,
			zmax=15,
			ztick={0, 3, 6, 9, 12},
			zlabel={$z$},
			view={-16.606849278776}{38.8085603165365},
			axis lines=box, 
			xmajorgrids,
			ymajorgrids,
			zmajorgrids,
			clip mode=individual,
			legend style={at={(0.5,0.9)}, anchor=north west, legend cell align=left, align=left}, 
			clip=false,
			grid style={dashed, gray!50, opacity=0.7} 
			]
			
			\addplot3 [ycomb, color=mycolor1, line width=2.0pt, mark size=5.0pt, mark=o, mark options={solid, mycolor1}]
			table[row sep=crcr] {%
				0 0 10\\
			};
			\addlegendentry{BS}
			
			\addplot3 [ycomb, color=mycolor2, line width=2.0pt, mark size=5.0pt, mark=+, mark options={solid, mycolor2}]
			table[row sep=crcr] {%
				10 10 15\\
			};
			\addlegendentry{IRS}
			
			\addplot3 [only marks, mark=*, mark options={}, mark size=2.5pt, color=mycolor3, fill=mycolor3] table[row sep=crcr]{%
				x y z\\
				55.5189072630706 4.65774589598349 1.5\\
				26.5668074706553 12.5172046599211 1.5\\
				21.4062014820729 18.8089432549816 1.5\\
				28.5184779958051 50.3746128821591 1.5\\
				63.7048608828261 10.3569683353091 1.5\\
				25.8702555567763 7.52990811093582 1.5\\
				63.5307164698516 16.3257118895927 1.5\\
				51.6935556196334 11.9989000053797 1.5\\
				46.03594757794 34.1732729154502 1.5\\
				34.9811754877588 30.6265271884563 1.5\\
			};
			\addlegendentry{Users}
			
		\end{axis}
	\end{tikzpicture}
    \caption{{Simulation setup for Fig.5 - Fig.11} }
    \label{sim_fig_1}
\end{figure}
The channel between the BS and RIS and the RIS and the $k$-th user are given by $\Vec{H}_1 = \sqrt{\frac{\mathrm{PL}_{\mathrm{LOS}}\left(d^{\mathrm{RB}}\right)}{ N}} \left(\sqrt{\frac{\kappa}{\kappa+1} } \bar{\Vec{H}}_1+\sqrt{\frac{1}{\kappa+1}} \tilde{\mathbf{H}}_1\right) $ and $\mathbf{h}_{2,k}^{\mathrm{u}}=\sqrt{\operatorname{PL}_{\mathrm{NLOS}}\left(d_{k}^{\mathrm{UR}}\right)} \tilde{\vec{h}}_{2,k}^{\mathrm{u}}$ respectively. Here, $\operatorname{PL}_{\mathrm{LOS}}\left(d\right)$ and $\operatorname{PL}_{\mathrm{NLOS}}\left(d\right)$ are the path losses experienced by the LOS and NLOS links at a distance of $d$ from the transmitter and are modelled using the 3GPP Urban
Micro (UMi) scenario from \cite[Table B.1.2.1-1]{3GPPTR36p814} under {$2.5$ GHz operating frequency.} Throughout the simulations, they are chosen as  $\operatorname{PL}_{\mathrm{LOS}}(d) = \frac{10^\frac{G_{t}+G_{r}-35.95}{10}}{d^{2.2}}$ and $\operatorname{PL}_{\mathrm{NLOS}}(d) = \frac{10^\frac{G_{t}+G_{r}-33.05}{10}}{d^{3.67}}$, where $G_t$ and $G_r$ denote the antenna gains (in dBi) at the transmitter and receiver respectively. It is assumed that the elements of BS have $5$ dBi gain while each user antenna and the RIS elements have $0$ dBi gain. Here, $\kappa$ is the Rician factor of the link between the BS and the RIS and is chosen to be $10$ throughout the simulations. The elements of the $M \times N$ matrix $\bar{\Vec{H}}_1$ are 
\begin{figure}[t!]
	\centering
	\includegraphics[scale=0.4,clip]{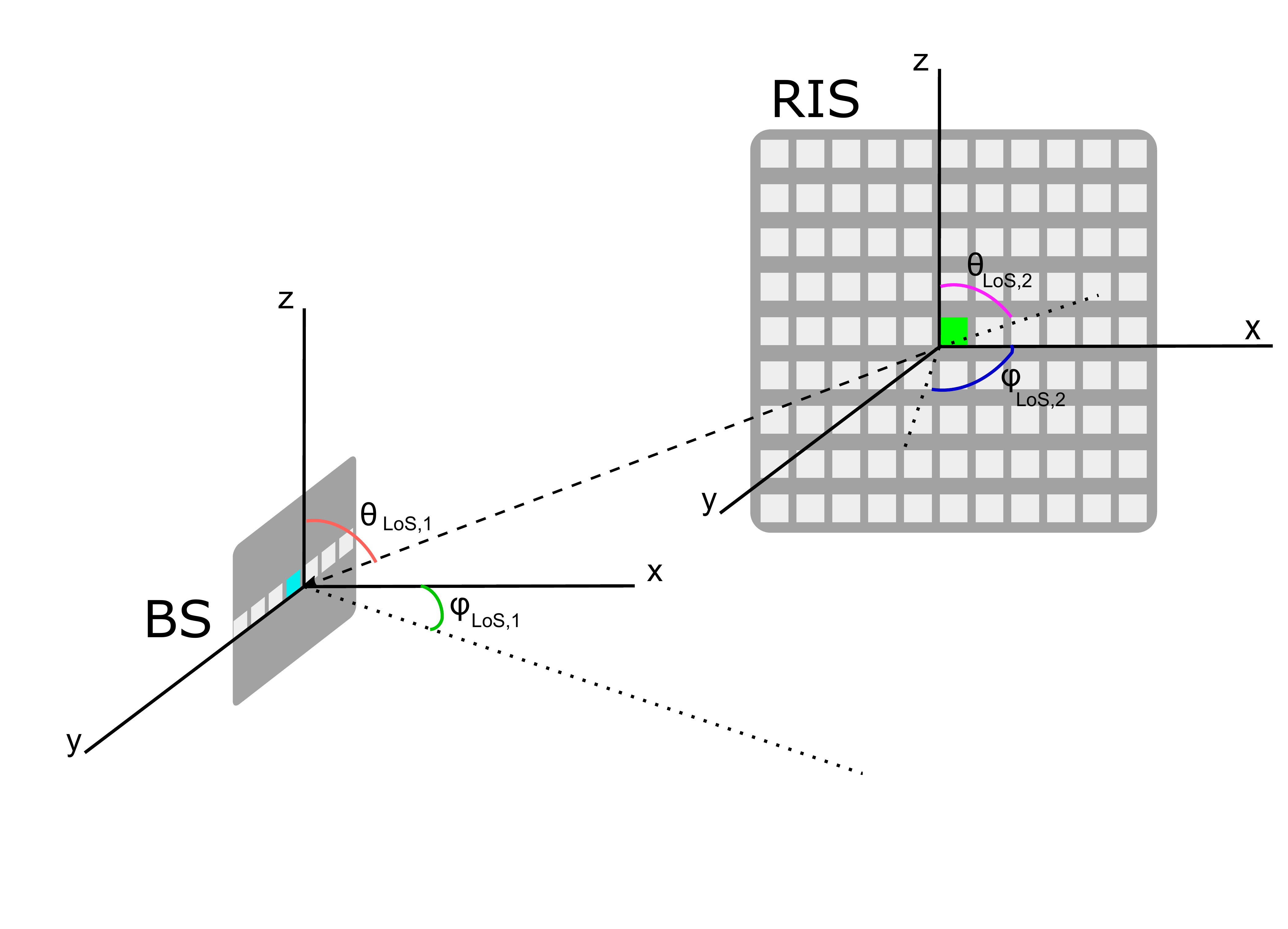}
	\caption{{Orientation of RIS with respect to BS}}
	\label{orientation}
\end{figure}

\begin{align}
& \left[\bar{\mathbf{H}}_{1}\right]_{m, n}\!\!\! \nonumber \\   & = \exp\! \Big[\frac{j2\pi}{\lambda} \left((m-1) d_{\text{BS}} \sin \left(\theta_{\text{LoS},1}(n)\right) \sin \left(\phi_{\text{LoS},1}(n)\right) \right. \nonumber \\ & \left.+ (n-1) d_{\text{RIS}} \sin \left(\theta_{\text{LoS},2}(m)\right) \sin \left(\phi_{\text{LoS},2}(m)\right)\right) \Big],
\end{align}
 where $\phi_{\text{LoS},1}$ and $\theta_{\text{LoS},1}$ represent the LoS azimuth and elevation angle of departures at the RIS and $\phi_{\text{LoS},2}$ and $\theta_{\text{LoS},2}$ represent the LoS azimuth and elevation angles of arrivals at the BS respectively. We show the orientation with respect to the centre of the BS and the RIS positions in Fig. \ref{orientation}. For the simulations,
$\theta_{\text{LoS},i}(n)$ and $\phi_{\text{LoS},i}(n)$ for $i \in \{1,2\}$
and $\forall n \in \{1,\cdots,N\}$, are generated uniformly between $0$ to $\pi$ and $0$ to $2\pi$ respectively, \cite{kammoun2020asymptotic}. Also, $d_{\text{BS}}$ and $d_{\text{RIS}}$ are the inter-antenna separation at the BS and the inter-element separation at the RIS and are chosen as $d_{\text{BS}}=d_{\text{RIS}}=0.5\lambda$. Each of the elements of $\tilde{\vec{H}}_1$ and $\tilde{\vec{h}}_{2,k}^u$ are standard normal RVs. Here, $\mathbf{\bar{H}}_1$ captures the orientation of the RIS with respect to the BS. Note that the proposed results are valid for all choices of $\mathbf{H}_1$ that satisfy the assumptions in Section III.B, and hence depending upon the position and the orientation of the RIS, appropriate models can be used to capture the effect on the system performance. Finally, $\Vec{H}_{2,k}$ is defined as $\Vec{H}_{2,k} =  \left[\mathbf{h}_{2,1}^{\mathrm{u}} \ \cdots \ \mathbf{h}_{2,K}^{\mathrm{u}} \right]$. Similarly, in the current set of simulations, we assume that all the users are data users and hence have $\text{SAR}_{\text{ref},k}=63 \times 10^{-4} \mathrm{~W} / \mathrm{Kg}$ per unit power and $\text{SAR}_{\text{max},k} =0.0029$ for all $ \forall k \in \mathcal{K}$. Similarly, the noise power in dBm is chosen as $\sigma^{2}=-174+10 \log _{10} B$ where the bandwidth for users using data is chosen as $100$ MHz. The maximum transmit power of each user ($p_{\text{max}}$) is chosen as $500$ mW. The correlation matrices, $\vec{R}_{\text{users}}$ and $\vec{R}_{\text{RIS}}$ are chosen according to the exponential correlation model, where the $(i,j)$-th matrix element is of the form $\eta^{|i-j|}$ for $\eta=0.95$.
\par Throughout the simulations, we compare the performance of the proposed optimization scheme with the results obtained using instantaneous CSI. Here, the BF and power allocation strategy discussed in sub-section \ref{bf_power_alloc} is used. The RIS phase optimization is performed using the scheme discussed in \cite{emfrisinst}, where a smooth approximation for the minimum is used, and projected gradient descent is used to find the optimal phase shift-vector. The different schemes compared in this paper are summarized in Table \ref{simu_schemes}. {Note that the proposed solution for phase shift design uses the deterministic equivalent of the minimum SINR, which holds when the problem dimensions tend to infinity with a fixed ratio, i.e. the number of antennas at the BS, the number of reflecting elements at the RIS and the number of users tend to infinity while having a fixed ratio among them. Using simulations, it was observed that this setting allows us to get a good  approximation for the minimum SINR even for finite system dimensions. Throughout the simulations, the parameters $K$, $M$, and $N$ are selected such that the system is operating in the deterministic equivalence regime.

\par Fig. \ref{asymp_montenum_65_minsinr}
and Fig. \ref{asymp_montenum_65_time1} compares the different schemes in terms of the minimum SINR and the time taken for evaluation, respectively. We can observe that the scheme which uses the asymptotic statistics for the phase optimization alone (Scheme 2) performs close to the scheme using the instantaneous channel gains (Scheme 1). Moreover, the time taken for the alternating optimization routine using the deterministic equivalent (Scheme 2) is considerably less than the scheme using the approximation for the minimum function (Scheme 1). We can also observe that there is a severe drop in the minimum SNR when both the power allocation and phase optimization routines are performed using the deterministic equivalent (Scheme 3). Thus, we can conclude that optimal power allocation is critical for system performance. However, the short computation time for the later scheme can be utilized to generate a good initial point for performing the phase optimization using other gradient-based schemes discussed in \cite{emfrisinst}, including the scheme using the approximation for the minimum. Also, note that the proposed solution for phase optimization needs to be re-evaluated only when the channel statistics vary. The rate at which the channel statistics vary will be very much less than the inverse of the coherence time of typical wireless channels. Hence, the RIS phase shifts using Scheme 1 need to be re-evaluated at a much higher frequency when compared to the solution using Scheme 2. Furthermore, the time taken for the evaluation of Scheme 1 is close to double the time taken for the evaluation of Scheme 2. This reduction in complexity is only possible due to the proposed phase optimization algorithm.  

\begin{figure}[t!]
\centering
%
%
\begin{tikzpicture}

\begin{axis}[%
width=0.95\linewidth,
at={(1.011in,0.642in)},
xmin=2,
xmax=10,
xlabel={K},
ymin=0,
ymax=0.4,
ylabel={Minimum SINR (dB)},
legend style={at={(0.03,0.97)}, anchor=north west, legend cell align=left, align=left},
grid
]
\addplot [color=blue, dashdotted, line width=2.0pt, mark=+, mark options={solid, blue}]
  table[row sep=crcr]{%
2	0.0792015895989806\\
4	0.150680354936343\\
6	0.229433122329713\\
10	0.361164261979047\\
};
\addlegendentry{Scheme 1}

\addplot [color=red, dashdotted, line width=2.0pt, mark=square, mark options={solid, red}]
  table[row sep=crcr]{%
2	0.0749354757196361\\
4	0.14063253751087\\
6	0.218\\
10	0.36084879808876\\
};
\addlegendentry{Scheme 2}

\addplot [color=black, dashdotted, line width=2.0pt, mark=asterisk, mark options={solid, black}]
  table[row sep=crcr]{%
2	0.0706025860521002\\
4	0.0760301417314828\\
6	0.112767634953176\\
10	0.179375182221765\\
};
\addlegendentry{Scheme 3}

\addplot [color=green, dashed, line width=2.0pt, mark=diamond, mark options={solid, green}]
  table[row sep=crcr]{%
2	0.0119670654103459\\
4	0.0200338029543605\\
6	0.0564475289388735\\
10	0.138111198767358\\
};
\addlegendentry{Scheme 6}

\end{axis}
\end{tikzpicture}%
\caption{Minimum SINR vs $K$ with $M=2\,K$  and $N=4 \,K$.}
\label{asymp_montenum_65_minsinr}
\end{figure}
\begin{figure}[t!]
\centering
%
%
\begin{tikzpicture}

\begin{axis}[%
width=0.95\linewidth,
at={(1.011in,0.642in)},
xmin=2,
xmax=10,
xlabel={K},
ymin=0,
ymax=35,
ylabel={Time (sec)},
legend style={at={(0.03,0.97)}, anchor=north west, legend cell align=left, align=left},
grid
]
\addplot [color=blue, dashdotted, line width=2.0pt, mark=+, mark options={solid, blue}]
  table[row sep=crcr]{%
2	2.567626051\\
4	10.390748503\\
6	23.245901311\\
10	31.705835635\\
};
\addlegendentry{Scheme 1}

\addplot [color=red, dashdotted, line width=2.0pt, mark=square, mark options={solid, red}]
  table[row sep=crcr]{%
2	1.273496095\\
4	4.206104146\\
6	12.818237665\\
10	14.903170009\\
};
\addlegendentry{Scheme 2}

\addplot [color=black, dashdotted, line width=2.0pt, mark=asterisk, mark options={solid, black}]
  table[row sep=crcr]{%
2	0.590547152999999\\
4	1.041812126\\
6	1.133294968\\
10	1.5\\
};
\addlegendentry{Scheme 3}

\end{axis}

\end{tikzpicture}%
\caption{{Time for evaluation vs $K$ with $M=2\,K$  and $N=4 \,K$.}}
\label{asymp_montenum_65_time1}
\end{figure}

\color{black}
{Next, in Fig.\ref{sinr_vs_power} we study the variation in minimum SINR with increasing values of the maximum transmit power} Note that the maximum allowable transmit power is limited by the exposure constraints, and hence beyond a specific point, increasing the transmit power will not increase the minimum SINR. 
\par We demonstrate the significance of the optimal phase shifts at the RIS elements in Fig.\ref{asymp_np_compare1}. Here, we compare the performance of Schemes 2 and 3 with the performance achieved by the same setup, except that the RIS is set to introduce no phase shift at the incoming wave. The figure shows that the performance degrades drastically without the phase shifts introduced by the RIS. Scheme 2 was consistently demonstrating superior performance with respect to Scheme 3; however, we can see that with the change in phase shifts, Scheme 5 achieves a very low minimum SINR. Moreover, without the right choice of phase shifts, we can see that Schemes 4 and 5 demonstrate almost similar performance. This further reiterates the significance of the right choice of phase shifts at the RIS to leverage the system utility. 

\begin{figure}
	\centering
%
%
\begin{tikzpicture}

\begin{axis}[%
width=0.95\linewidth,
at={(1.011in,0.667in)},
xmin=0.1,
xmax=0.25,
xlabel={$p_{\rm max}$ (mW)},
ymin=0.2,
ymax=1,
ylabel={Minimum SINR (dB)},
yticklabel style={yshift=4pt},
legend style={at={(0.03,0.99)}, anchor=north west, legend cell align=left, align=left},
grid,
xtick={0.1,0.15,0.2,0.25}, 
]
\addplot [color=blue, dashdotted, line width=2.0pt, mark=+, mark options={solid, blue}]
  table[row sep=crcr]{%
0.1	0.627819254698885\\
0.15	0.760479198363625\\
0.2	0.815498509497439\\
0.25	0.815498509497439\\
};
\addlegendentry{Scheme 1}

\addplot [color=red, dashdotted, line width=2.0pt, mark=square, mark options={solid, red}]
  table[row sep=crcr]{%
0.1	0.636271822681461\\
0.15	0.774385135853335\\
0.2	0.8285739276804\\
0.25	0.8285739276804\\
};
\addlegendentry{Scheme 2}

\addplot [color=black, dashdotted, line width=2.0pt, mark=asterisk, mark options={solid, black}]
  table[row sep=crcr]{%
0.1	0.369320499504431\\
0.15	0.447021301434778\\
0.2	0.476289053863139\\
0.25	0.476289053863139\\
};
\addlegendentry{Scheme 3}

\end{axis}

\end{tikzpicture}%
	\caption{{Minimum SINR vs $p_{\text{max}}$}, for $K=4$, $M=8$, and $N=25$.}
	\label{sinr_vs_power}
\end{figure}

\begin{figure}
\centering
%
%
\begin{tikzpicture}

\begin{axis}[%
width=0.95\linewidth,
at={(1.011in,0.642in)},
xmin=2,
xmax=10,
xlabel={$K$},
ymin=0,
ymax=0.4,
ylabel={Minimum SINR (dB)},
legend style={at={(0.03,0.97)}, anchor=north west, legend cell align=left, align=left},
grid
]
\addplot [color=blue, dashdotted, line width=2.0pt, mark=+, mark options={solid, blue}]
  table[row sep=crcr]{%
2	0.0576840181338714\\
4	0.131286187154506\\
6	0.218\\
10	0.361164261979047\\
};
\addlegendentry{Scheme 2}

\addplot [color=red, dashdotted, line width=2.0pt, mark=square, mark options={solid, red}]
  table[row sep=crcr]{%
2	0.0500000000000007\\
4	0.0760301417314828\\
6	0.112767634953176\\
10	0.179375182221765\\
};
\addlegendentry{Scheme 3}

\addplot [color=black, dashdotted, line width=2.0pt, mark=asterisk, mark options={solid, black}]
  table[row sep=crcr]{%
2	0.0123197431898276\\
4	0.0181685777501119\\
6	0.0247892821098574\\
10	0.025652684104946\\
};
\addlegendentry{Scheme 4}

\addplot [color=green, dashed, line width=2.0pt, mark=diamond, mark options={solid, green}]
  table[row sep=crcr]{%
2	0.012583618702946\\
4	0.019454865530335\\
6	0.0231247969274033\\
10	0.0257009452239263\\
};
\addlegendentry{Scheme 5}

\end{axis}

\end{tikzpicture}%
\caption{Minimum SINR vs $K$ with $M=2\,K$  and $N=4 \,K$.}
\label{asymp_np_compare1}
\end{figure}

\begin{figure}
%
%
\begin{tikzpicture}

\begin{axis}[%
width=0.95\linewidth,
at={(1.011in,0.642in)},
xmin=10,
xmax=60,
xlabel={$N$},
ymin=0,
ymax=0.8,
ylabel={Minimum SINR (dB)},
legend style={at={(0.03,0.97)}, anchor=north west, legend cell align=left, align=left},
grid
]
\addplot [color=blue, dashdotted, line width=2.0pt, mark=+, mark options={solid, blue}]
  table[row sep=crcr]{%
10	0.0900830361607774\\
20	0.212189266013596\\
30	0.321466808771198\\
40	0.481151682209308\\
50	0.678617232722139\\
60	0.770220924839315\\
};
\addlegendentry{Scheme 1}

\addplot [color=red, dashdotted, line width=2.0pt, mark=square, mark options={solid, red}]
  table[row sep=crcr]{%
10	0.0662523810274678\\
20	0.203060864529029\\
30	0.307357258510471\\
40	0.494488847544638\\
50	0.679756609823556\\
60	0.780476366419975\\
};
\addlegendentry{Scheme 2}

\addplot [color=black, dashdotted, line width=2.0pt, mark=asterisk, mark options={solid, black}]
  table[row sep=crcr]{%
10	0.0387238048888747\\
20	0.106283370179014\\
30	0.161072293563997\\
40	0.274046208842989\\
50	0.403536325172787\\
60	0.482869811112003\\
};
\addlegendentry{Scheme 3}

\end{axis}

\end{tikzpicture}%
	\caption{{Minimum SINR vs $N$, for $K=4$ and $M=8$.}}
	\label{N_vary1}
\end{figure}

\begin{figure}
   \centering
%
%
\begin{tikzpicture}

\begin{axis}[%
width=0.95\linewidth,
at={(1.011in,0.642in)},
xmin=2,
xmax=10,
xlabel={$K$},
ymin=0,
ymax=0.4,
ylabel={Minimum SINR (dB)},
legend style={at={(0.03,0.97)}, anchor=north west, legend cell align=left, align=left},
grid
]
\addplot [color=blue, dashdotted, line width=2.0pt, mark=+, mark options={solid, blue}]
  table[row sep=crcr]{%
2	0.0336617518526392\\
4	0.0906154763131468\\
6	0.117045642231119\\
10	0.281740398922347\\
};
\addlegendentry{$\text{SAR}_{\text{max}}=0.001$}

\addplot [color=red, dashdotted, line width=2.0pt, mark=square, mark options={solid, red}]
  table[row sep=crcr]{%
2	0.0653829380845465\\
4	0.144963285017552\\
6	0.206353348069841\\
10	0.363299579324782\\
};
\addlegendentry{$\text{SAR}_{\text{max}}=0.003$}

\addplot [color=black, dashdotted, line width=2.0pt, mark=asterisk, mark options={solid, black}]
  table[row sep=crcr]{%
2	0.0792000000000002\\
4	0.150700000000001\\
6	0.2294\\
10	0.3612\\
};
\addlegendentry{$\text{SAR}_{\text{max}}=0.004$}

\end{axis}
\end{tikzpicture}%
\caption{Minimum SINR vs $K$ for different values of $\text{SAR}_{\text{max},k}$ with $M=2\,K$  and $N=4 \,K$.}
\label{vary_sar_max}
\end{figure}

In Fig. \ref{N_vary1}, we demonstrate the variation in minimum SINR with an increasing number of RIS elements. As expected, the minimum SINR increases with increasing $N$. This further suggests that for a system with a fixed number of users and a given number of antennas at the BS, we can improve the system performance by using a RIS with more reflecting elements. Furthermore, we can see that Scheme 2 can achieve almost the same performance achieved using Scheme 1. This means using the asymptotic statistics also we can achieve what we were achieving using the instantaneous CSI. Scheme 2 needs much fewer computations, and feedback overhead compared to the methods using instantaneous CSI is also to be noted. 
{We also study the effect of the maximum allowable exposure for each user on the minimum SINR achieved. In Fig. \ref{vary_sar_max}, we plot the variation in minimum SINR with respect to the increasing number of users for different values of $\text{SAR}_{\max}$. As expected, we can see that the minimum SINR increases as the allowed exposure limit increases. This is intuitive since an increase in the maximum allowed exposure would increase the maximum power that can be used for transmission, and hence it improves the SINR for each user. Also, increasing the maximum exposure beyond a value does not improve the minimum SINR since we are also restricted by the transmit power constraints.}
\par Note that the proposed results are valid for all choices of $\mathbf{H}_1$ that satisfy Assumption 2 in Section III.B. To demonstrate the general nature of the results proposed, we also study the performance for a scenario where the LoS component of the link between the BS and the RIS is modelled using \cite[(4)]{rinchi2022compressive}, for the number of scatterers, $L=M$. For this simulation, the users are uniformly spread between the area of overlap between the first quadrant and two circles of radius $R_{min} = 10$ m and $R_{max} = 70$ m centered at the RIS. Fig. 10, shows one sample realization of the simulation setup. The minimum SINR for different numbers of users in this setup is given in Fig. \ref{new_H_sim_res}. Similar to the results observed for the other channel model, here also we can see that Scheme 1 and Scheme 2 achieve very close performance.

\begin{figure}
 \centering
%
%
\definecolor{mycolor1}{rgb}{0.00000,0.44700,0.74100}%
\definecolor{mycolor2}{rgb}{0.85000,0.32500,0.09800}%
\definecolor{mycolor3}{rgb}{0.92900,0.69400,0.12500}%
\begin{tikzpicture}

\begin{axis}[%
width=0.95\linewidth, 
height=0.75\linewidth, 
at={(0.844in,0.642in)},
xmin=0,
xmax=15, 
xtick={ 0, 5,10, 15},
tick align=outside,
xlabel={$x$},
ymin=0,
ymax=10,
ytick={0, 5,10, 15},
ylabel={$y$},
zmin=0,
zmax=15,
ztick={ 0,  3,  6,  9, 12, 15},
zlabel={$z$},
view={-43.6150684931507}{28.3984435797665},
axis x line*=bottom,
axis y line*=left,
axis z line*=left,
clip mode=individual,
axis lines=box, 
xmajorgrids,
ymajorgrids,
zmajorgrids,
legend style={at={(0.5,0.85)}, anchor=north west, legend cell align=left, align=left},
grid style={dashed, gray!50, opacity=0.7} 
]
\addplot3 [ycomb, color=mycolor1, line width=2.0pt, mark size=5.0pt, mark=o, mark options={solid, mycolor1}]
 table[row sep=crcr] {%
0	0	10\\
};
 \addlegendentry{BS}

\addplot3 [ycomb, color=mycolor2, line width=2.0pt, mark size=5.0pt, mark=+, mark options={solid, mycolor2}]
 table[row sep=crcr] {%
10	10	15\\
};
 \addlegendentry{IRS}

\addplot3[only marks, mark=*, mark options={}, mark size=1.5000pt, color=mycolor3, fill=mycolor3] table[row sep=crcr]{%
x	y	z\\
13.9012021461291	9.60717251301272	1.5\\
13.9656362330765	5.33269613542739	1.5\\
11.8976291050468	5.89408604713722	1.5\\
11.5290143915766	6.3434959467844	1.5\\
12.0370341425575	8.59818663443994	1.5\\
10.352083136678	5.29781077572968	1.5\\
14.5503472059161	5.73978345252208	1.5\\
11.8478753969126	5.53785057935256	1.5\\
14.5379083192751	6.16612227782805	1.5\\
13.6923968299738	5.85706428609855	1.5\\
};
\addlegendentry{Users}

\end{axis}
\end{tikzpicture}%
\caption{{Simulation setup for Fig. \ref{new_H_sim_res}}}
\label{simu_setup_2}
\end{figure}

\begin{figure}
\centering
%
%
\begin{tikzpicture}

\begin{axis}[%
width=0.95\linewidth,
at={(1.011in,0.642in)},
xmin=2,
xmax=10,
xlabel={$K$},
ymin=0,
ymax=0.12,
ylabel={Minimum SINR (dB)},
legend style={at={(0.03,0.97)}, anchor=north west, legend cell align=left, align=left},
grid,
y tick label style={/pgf/number format/.cd,fixed,precision=3}
]
\addplot [color=blue, dashdotted, line width=2.0pt, mark=+, mark options={solid, blue}]
  table[row sep=crcr]{%
2	0.00131920213760139\\
4	0.022424127251897\\
6	0.0592317738028019\\
10	0.100112536000248\\
};
\addlegendentry{Scheme 1}

\addplot [color=red, dashdotted, line width=2.0pt, mark=square, mark options={solid, red}]
  table[row sep=crcr]{%
2	0.00627094833000896\\
4	0.0223277516202067\\
6	0.0549999999999997\\
10	0.0992681693256827\\
};
\addlegendentry{Scheme 2}

\addplot [color=black, dashdotted, line width=2.0pt, mark=asterisk, mark options={solid, black}]
  table[row sep=crcr]{%
2	0.00130809045122682\\
4	0.0136688409095065\\
6	0.0238551703079093\\
10	0.0356286315621048\\
};
\addlegendentry{Scheme 3}

\end{axis}

\end{tikzpicture}%
\caption{{Minimum SINR vs $K$ with $M=2\,K$  and $N=4 \,K$.}}
\label{new_H_sim_res}
\end{figure}

\section{Conclusion}\label{conclusion}
We consider the uplink communication of a RIS-assisted system and study the joint optimization of the BF vectors, power allocation, and the phase shifts at the RIS elements. A simple projected gradient descent-based optimization strategy is proposed to identify the optimal phase shift vector at the RIS elements that maximize the minimum SINR subject to constraints on the power consumed by the users. The proposed solution depends only on the asymptotic channel statistics and hence needs to be updated only when the large-scale channel statistics change. The simulations demonstrate that the minimum SINR achieved using the solution is very close to the minimum SINR achieved using the phase vector designed using the instantaneous channel statistics. 
Furthermore, the numerical results show that the proposed algorithms achieve more than $100 \%$ gain in terms of minimum SINR, compared to a system with random RIS phase shifts, with $40$ RIS elements, $20$ antennas at the BS, and $10$ users.

\begin{appendices}
	\section{Some useful results}
	
	\begin{lemma}\label{Quadratic}
		(Convergence of Quadratic Forms): Let $\mathbf{y} \sim \mathcal{C} \mathcal{N}\left(\mathbf{0}_{M}, \mathbf{I}_{M}\right) .$ Let $\mathbf{A}$ be an $M \times M$ matrix independent of $\mathbf{y}$, which has a bounded spectral norm. Then {$\frac{1}{M} \mathbf{y}^{H} \mathbf{A} \mathbf{y}-\frac{1}{M} \operatorname{tr}(\mathbf{A}) \xrightarrow[M\to \infty]{\text{a.s.}} 0$}.
	\end{lemma}
\begin{lemma}
	Consider the $M\times M$ matrix:
	$$
	{\bf B}=\frac{1}{K}{\bf U}{\bf X}{\bf X}^{H}{\bf U}^{H},
	$$
	with the following Assumptions:
	\begin{enumerate}
		\item ${\bf X}\in\mathbb{C}^{N\times K}$ is such that $[{\bf X}]_{ij}$ are identically distributed and  independent and ${\rm var}([{\bf X}]_{ij})=1$ and finite fourth order moments.
		\item ${\bf R}={\bf U}{\bf U}^{H}$ with ${\bf U}\in \mathbb{C}^{M\times N} $ has uniformly bounded spectral norm over $M$ and $N$. 
		\item Denoting $c_1=\frac{K}{M}$ and $c_0=\frac{N}{K}$, then
		$$
		0< \liminf c_1<\infty \ \ \text{and} \ \ 0< c_0<\infty. 
		$$
		Then, as all $M, K$ and $N$ grow large with ratios $c_1$ and $c_0$, for $z\in\mathbb{C}\backslash\mathbb{R}^{+}$,  and for all non-negative hermitian matrices ${\bf C}\in\mathbb{C}^{N\times N}$ with uniformly bounded spectral norm: $\frac{1}{K}\mathrm{Trace}  \left\{{\bf C}\left({\bf B}-z{\bf I}\right)^{-1}\right\}-\frac{1}{K}\mathrm{Trace}\left\{ {\bf C}\left(-{z}{\bf I}+\frac{{\bf R}}{1+e(z)}\right)^{-1}\right\}$
converges almost surely to zero where $e(z)$ form the unique solution of
\begin{equation}
e(z)=\frac{1}{K}\mathrm{Trace}\left\{ {\bf R}\left(-{z}{\bf I}+\frac{{\bf R}}{1+e(z)}\right)^{-1}\right\}. \nonumber 
\end{equation}
	\end{enumerate}
\label{lem:stiel}
\end{lemma}

\section{Proof of Theorem \ref{th:1}}
\label{app:A}

	We start by proving that
	\begin{align}
&	d= \nonumber \\ & \frac{1}{K}\text{Trace} \left \lbrace  \vec{U}\vec{U}^H \left(\vec{U}\vec{U}^H  \frac{\tau}{{d}(1+\tau)}+ {\sigma}^2 \mathbf{I}_{M} \right)^{-1} \right \rbrace, \label{eq:dd}
	\end{align}
	admits a unique solution $\overline{d}(\tau)$.
Dividing both sides in \eqref{eq:dd} by $\overline{d}(\tau)$, we obtain
	 \begin{align}
	 & K= \nonumber \\ & \text{Trace}  \left \lbrace  \vec{U}\vec{U}^H \left(\vec{U}\vec{U}^H  \frac{\tau}{(1+\tau)}+ {\sigma}^2 \overline{d}(\tau)\mathbf{I}_{M} \right)^{-1} \right \rbrace. \label{eq:fixed}
	 \end{align}
	The function \\ $h_\tau:d\mapsto \frac{1}{K}\text{Trace}  \left \lbrace  \vec{U}\vec{U}^H \left(\vec{U}\vec{U}^H  \frac{\tau}{(1+\tau)}+ {\sigma}^2 d\mathbf{I}_{M} \right)^{-1} \right \rbrace$, is decreasing. Since $K/M<1$, $\lim_{d\to 0} h_\tau(d)>1$. As $\lim_{d\to \infty} h_\tau(d)=0$, there exists a unique $\overline{d}(\tau)$ satisfying $h_\tau(\overline{d}(\tau))=1$, which shows existence and uniqueness of the solution to \eqref{eq:dd}. On the other hand, since $\lim_{d\to \infty}\sup_{\tau\geq 0} h_\tau(d) =0$, there exists $C$ such that $\sup_{\tau\geq 0}\overline{d}(\tau)<C$. Now, since $h(0)\geq\frac{1+\tau}{\tau}\liminf \frac{M}{K}>1$, we also have $\inf_{\tau\geq 0}\overline{d}(\tau)>0$.\\
The next step is now to show that:
$$
\sup_{\tau\in\mathcal{I}}\max_{1\leq k \leq K} |\frac{d_k(\tau)}{\overline{d}(\tau)}-1|\underset{M \rightarrow \infty}{\longrightarrow} 0,
$$
where $d_1(\tau),\cdots,d_K(\tau)$ are the unique solutions to the system of equations in \eqref{eq:r}. Letting $e_k(\tau):=\frac{d_k(\tau)}{\overline{d}(\tau)}$, this amounts to showing that for all $\ell>0$, we have for sufficiently large dimensions:
\begin{equation}
\forall \tau\in\mathcal{I}, \ \  k \in \mathcal{K}, \ \  -\ell\leq e_k(\tau)-1\leq \ell. \label{eq:required} 
\end{equation}
Assume without loss of generality that $e_1(\tau)<e_2(\tau)<\cdots<e_K(\tau)$. Proving \eqref{eq:required} is thus equivalent to showing the following inequalities:
\begin{align}
\forall \tau\in\mathcal{I}, \ \ e_K(\tau)\leq 1+\ell, \label{eq:upbound}\\
\forall \tau\in\mathcal{I}, \ \	e_1(\tau)\geq 1-\ell. \label{eq:lower_bound}
\end{align}
\noindent{\underline{Proof of \eqref{eq:upbound}}.} Starting  from the expression of $d_k(\tau)$, we may express $e_k(\tau)$ as
\begin{align}
& e_k(\tau) = \frac{\tilde{\vec{h}}_{2,k}^H \vec{U}^H}{K\overline{d}(\tau)} \times  \nonumber \\ 
& \left( \sum\limits_{i\neq k} \frac{\tau}{K d_i(\tau)} \vec{U}\tilde{\vec{h}}_{2,i} \tilde{\vec{h}}_{2,i}^H \vec{U}^H  + {\sigma}^2 \mathbf{I}_{M} \right)^{-1} \vec{U} \tilde{\vec{h}}_{2,k}, \nonumber \\
 & = \frac{\tilde{\vec{h}}_{2,k}^H \vec{U}^H}{K} \times \nonumber \\ &  \left( \sum\limits_{i\neq k} \frac{\tau \vec{U}\tilde{\vec{h}}_{2,i} \tilde{\vec{h}}_{2,i}^H \vec{U}^H}{K  e_i(\tau) }   + {\sigma}^2 \bar{d}(\tau) \mathbf{I}_{M} \right)^{-1} \vec{U} \tilde{\vec{h}}_{2,k} \label{e_k_1}  .
\end{align}
 Evaluating \eqref{e_k_1} for $k=K$, we obtain 
 \begin{align}
& e_K(\tau)= \frac{\tilde{\vec{h}}_{2,K}^H \vec{U}^H}{K} \times \nonumber \\ & \left(\! \sum\limits_{i\neq K} \frac{\tau}{K e_i(\tau) } \vec{U}\tilde{\vec{h}}_{2,i} \tilde{\vec{h}}_{2,i}^H \vec{U}^H \! \!+\! {\sigma}^2 \bar{d}(\tau) \mathbf{I}_{M} \right)^{\!-1}\!\!\!\!\!\!\!\! \vec{U} \tilde{\vec{h}}_{2,K}. \nonumber 
 \end{align}

Since $e_K(\tau)\geq e_i(\tau)$ for all $i\neq K$, we have 
\begin{align}
& e_K(\tau)\leq \frac{1}{K} \tilde{\vec{h}}_{2,K}^H \vec{U}^H \times \nonumber \\ &  \left( \sum\limits_{i\neq K} \frac{\tau  \vec{U}\tilde{\vec{h}}_{2,i} \tilde{\vec{h}}_{2,i}^H \vec{U}^H }{K e_K(\tau) } + {\sigma}^2 \bar{d}(\tau) \mathbf{I}_{M} \right)^{-1} \vec{U} \tilde{\vec{h}}_{2,K},
\end{align}
or equivalently 
\begin{align}
& 1\leq  \frac{1}{K} \tilde{\vec{h}}_{2,K}^H \vec{U}^H \nonumber \\
 & \left( \sum\limits_{i\neq K} \frac{\tau \vec{U}\tilde{\vec{h}}_{2,i} \tilde{\vec{h}}_{2,i}^H \vec{U}^H}{K  }   + {\sigma}^2 \bar{d}(\tau)e_K(\tau) \mathbf{I}_{M} \right)^{-1} \vec{U} \tilde{\vec{h}}_{2,K}. \label{eq:prev}
\end{align}
Recall that our objective is to prove that for sufficiently large dimensions:
$$
\sup_{\tau \in\mathcal{I}}e_K(\tau)\leq 1+\ell,
$$
or equivalently:
$$
\limsup_{N,M,K} \sup_{\tau \in\mathcal{I}} e_K(\tau) \leq 1+\ell.
$$
For that, we proceed by contradiction and assume that:
\begin{equation}
\limsup_{N,M,K} \sup_{\tau\in\mathcal{I}} e_K(\tau) \geq 1+\ell . \label{eq:nec}
\end{equation}
Hence, in view of \eqref{eq:prev}, for $N,K$ and $M$ sufficiently large, there exists $\tilde{\tau}\in\mathcal{I}$ such that
\begin{align}
& 1\leq \frac{1}{K} \tilde{\vec{h}}_{2,K}^H \vec{U}^H \times  \nonumber \\ &  \left( \sum\limits_{i\neq K} \frac{\tilde{\tau} \vec{U}\tilde{\vec{h}}_{2,i} \tilde{\vec{h}}_{2,i}^H \vec{U}^H}{K  }   + {\sigma}^2 \bar{d}(\tilde{\tau})(1+\ell) \mathbf{I}_{M} \right)^{-1} \vec{U} \tilde{\vec{h}}_{2,K}. \nonumber
\end{align}
 
To continue, we apply Lemma \ref{Quadratic} and Lemma \ref{lem:stiel} to study the asymptotic limit of the right-hand side term of the above inequality. Indeed, the following convergence holds point-wise in $\tau$: 
\begin{align}
 \frac{1}{K} \tilde{\vec{h}}_{2,K}^H \vec{U}^H &  \left( \sum\limits_{i\neq K} \frac{\tau}{K  } \vec{U}\tilde{\vec{h}}_{2,i} \tilde{\vec{h}}_{2,i}^H \vec{U}^H  + {\sigma}^2 \bar{d}(\tau)(1+\ell) \mathbf{I}_{M} \right)^{-1} \nonumber  \\ & \vec{U} \tilde{\vec{h}}_{2,K}-\mu  \xrightarrow[M\to \infty]{\text{a.s.}} 0, \label{eq:conv}
\end{align}
where $\mu(\tau)$ is the unique solution to the following equation:
$$
\mu(\tau)=\frac{1}{K}\text{Trace}\left\{ {\bf U}{\bf U}^{H}\left(\sigma^2\overline{d}(\tau)(1+\ell){\bf I}+\frac{\tau{\bf U}{\bf U}^{H}}{1+\mu(\tau)\tau}\right)^{-1}\right\}
$$
$\sup_{\tau\in\mathcal{I}} \left|\frac{\tilde{\vec{h}}_{2,K}^H \vec{U}^H}{K} \left( \sum\limits_{i\neq K} \frac{\tau}{K  } \vec{U}\tilde{\vec{h}}_{2,i} \tilde{\vec{h}}_{2,i}^H \vec{U}^H  + {\sigma}^2 \bar{d}(\tau)(1+\ell) \mathbf{I}_{M} \right)^{-1} \right.$ $\left.  \vec{U} \tilde{\vec{h}}_{2,K}-\mu \right|\underset{M \rightarrow \infty}{\longrightarrow} 0$.
Since function $(\tau,d)\mapsto $ $\frac{1}{K} \tilde{\vec{h}}_{2,K}^H \vec{U}^H  \times $ $ \left( \sum\limits_{i\neq K} \frac{\tau}{K  } \vec{U}\tilde{\vec{h}}_{2,i} \tilde{\vec{h}}_{2,i}^H \vec{U}^H  + {\sigma}^2 \bar{d}(\tau)(1+\ell) \mathbf{I}_{M} \right)^{-1}  $ $  \vec{U} \tilde{\vec{h}}_{2,K}$ is Lipschitz on any compact set in $(0,\infty)\times(0,\infty)$, by Arzela-Ascoli theorem, the convergence in \eqref{eq:conv} hold uniformly in $\tau$ and $\overline{d}(\tau)$ belonging to compact sets in $(0,\infty)\times (0,\infty) $. As we proved earlier that $\overline{d}(\tau)$ lies in a compact set when $\tau\in\mathcal{I}$, we obtain:

In other words, for any $\epsilon$ sufficiently small and positive, and sufficiently large $N, K$ and $M$, $\forall \ \ \tau\in\mathcal{I},  $
\begin{align}
& \mu(\tau)+\epsilon \geq \frac{\tilde{\vec{h}}_{2,K}^H \vec{U}^H}{K} \times \nonumber \\ &  \left( \sum\limits_{i\neq K} \frac{\tau}{K  } \vec{U}\tilde{\vec{h}}_{2,i} \tilde{\vec{h}}_{2,i}^H \vec{U}^H  + {\sigma}^2 \bar{d}(\tau)(1+\ell) \mathbf{I}_{M} \right)^{-1} \vec{U} \tilde{\vec{h}}_{2,K}. \nonumber 
\end{align}

Particularly, for $\tau=\tilde{\tau}$, we obtain:
\begin{align}
\frac{\tilde{\vec{h}}_{2,K}^H \vec{U}^H }{K} & \left( \sum\limits_{i\neq K} \frac{\tilde{\tau}}{K  } \vec{U}\tilde{\vec{h}}_{2,i} \tilde{\vec{h}}_{2,i}^H \vec{U}^H  + {\sigma}^2 \bar{d}(\tilde{\tau})(1+\ell) \mathbf{I}_{M} \right)^{-1} \nonumber \\ & \vec{U} \tilde{\vec{h}}_{2,K}\leq \mu(\tilde{\tau})+\epsilon. \label{eq:above}
\end{align}
To show contradiction to \eqref{eq:nec}, it suffices to prove that for all $\tau\in\mathcal{I}$, $\mu(\tau)<1$. Indeed, if $\forall \tau\in\mathcal{I}, \ \  \mu(\tau)<1$, then by choosing $\epsilon<\ell$, the inequality in \eqref{eq:above} becomes in contradiction to \eqref{eq:nec}. Recalling that $\mu(\tau)$ is the unique solution to the following equation:
$$
\mu(\tau)=\frac{1}{K}\text{Trace}\left\{ {\bf U}{\bf U}^{H}\left(\sigma^2\overline{d}(\tau)(1+\ell){\bf I}+\frac{\tau{\bf U}{\bf U}^{H}}{1+\mu(\tau)\tau}\right)^{-1}\right\}.
$$
Dividing both sides by $\mu(\tau)$, we obtain
\begin{align}
 K & = \nonumber \\ &  \text{Trace}\left\{ {\bf U}{\bf U}^{H}\left({\sigma^2}\mu(\tau)\overline{d}(\tau)(1+\ell){\bf I} \right. \right. \nonumber \\ & \left. \left. +\frac{\mu(\tau)\tau{\bf U}{\bf U}^{H}}{1+\mu(\tau)\tau}\right)^{-1}\right\}.  \label{eq:rf}  
\end{align}

Let \\ $g:\mu \mapsto \frac{1}{K}\text{Trace}\left\{ {\bf U}{\bf U}^{H}\left({\sigma^2}\mu\overline{d}(\tau)(1+\ell){\bf I}+\frac{\mu\tau{\bf U}{\bf U}^{H}}{1+\mu\tau}\right)^{-1}\right\}$. Then, $g$ is a decreasing function of $\mu$. Moreover, 
\begin{align}
& g(1)=  \nonumber \\ & \frac{1}{K}\text{Trace}\left\{ {\bf U}{\bf U}^{H}\left({\sigma^2}\overline{d}(\tau)(1+\ell){\bf I}+\frac{\tau{\bf U}{\bf U}^{H}}{1+\tau}\right)^{-1}\right\}, \\
&< \frac{1}{K}\text{Trace}\left\{ {\bf U}{\bf U}^{H}\left({\sigma^2}\overline{d}(\tau){\bf I}+\frac{\tau{\bf U}{\bf U}^{H}}{1+\tau}\right)^{-1}\right\}=1,
\end{align}
where in the last equality, we used \eqref{eq:fixed}. If $\mu(\tau)\geq 1$, then since $g$ is decreasing $g(\mu(\tau))\leq g(1)<1$, creating a contradiction with \eqref{eq:rf}. Hence, $\mu(\tau)<1$. This completes the proof of \eqref{eq:upbound}. 
The proof of \eqref{eq:lower_bound} follows the same lines and is thus omitted. 
\section{Proof of Theorem \ref{th:th_2}}
\label{app:B}

\noindent{\bf Strategy of the proof.} 
We organize the proof of Theorem \ref{th:th_2} in two parts. In the first part, denoting by $\alpha_k=K\ell_kp_{{\rm max},k}, \forall  k \in \mathcal{K}$, we prove that the optimal cost of the following problem:

\begin{equation} 
\tag{$\hat{\mathcal{P}}$}
\begin{aligned}
&  \tau^*=	\displaystyle{\max_{d_k,\tau}}   \tau \nonumber  \\
&  \text { s.t. }  \ \ d_k(\tau)=h_k(d_1,\cdots,d_K;\tau), \ \ k \in \mathcal{K} \nonumber \\
&  \frac{\tau}{d_k(\tau)}\leq \alpha_k,  \ \ \forall  k \in \mathcal{K}.
\end{aligned}
\end{equation}

is asymptotically equivalent to the optimal cost of the following deterministic problem:
\begin{equation}
\tag{$\mathcal{P}_{\rm asym}$}
 \begin{aligned}
&	\tilde{\tau}=	\displaystyle{\max_{{d},\tau}} \ \  \tau  \\
&  \text { s.t. } \ \ {d}=\frac{1}{K}\text{Trace}  \left \lbrace  \vec{U}\vec{U}^H \left(\vec{U}\vec{U}^H  \frac{\tau}{{{d}}(1+\tau)}+ {\sigma}^2 \mathbf{I}_{M} \right)^{-1} \right \rbrace, \\
	&  \ \ \frac{\tau}{{d}}\leq \alpha_k , \ \ \forall  k \in \mathcal{K}.
\end{aligned}
\end{equation}
in the sense that $\tau^\star-\tilde{\tau}\underset{M \rightarrow \infty}{\longrightarrow} 0$. 
In the second part, we show that $\tilde{\tau}$ is equal to $\overline{\tau}$, the unique solution to \eqref{eq:tau_bar}.  Finally, by combining the findings of Theorem \ref{th:1} along with those of both parts, we deduce the convergence results of Theorem \ref{th:th_2}.\\
\noindent{\bf First part.}
Let $(\tilde{\tau},\breve{d})$ be the solution to $(\mathcal{P}_{\rm asym})$.  The uniqueness of this solution is a straightforward consequence of the second part of the proof and thus will be admitted here. Our goal in this section is to show that $\tau^*-\tilde{\tau}\underset{M \rightarrow \infty}{\longrightarrow} 0$. 
 \\
 Before delving into the proof, we need first to check that the solution $\tau^*$ to $\hat{\mathcal{P}}$ lies almost surely in a compact set in $(0,\infty)$ to enable the use of the results of Theorem \ref{th:1}. For that, we use the fact that for any feasible $\tau,d_k$ such $\frac{\tau}{d_k}\leq \alpha_k$, we have:
 \begin{align}
  & \frac{1}{K}\tilde{\bf g}_k^{H}\left(\sum_{i\neq k}\alpha_i\tilde{\bf g}_i\tilde{\bf g}_i^{H}+\sigma^2{\bf I} \right)^{-1}\tilde{\bf g}_k \nonumber \\ & \leq d_k=h_k(d_1,\cdots,d_K;\tau)\leq \frac{1}{K\sigma^2}\tilde{\bf g}_k^{H}{\bf g}_k.
 \end{align}
 From standard results of random matrix theory, under the regime growth of Assumption 1 and Assumption 2, there exists strictly positive constants $C_1$ and $C_2$ such that almost surely,
 \begin{align}
 \forall k=1,\cdots, K, \ \ \ \frac{1}{K}\tilde{\bf g}_k^{H}\left(\sum_{i\neq k}\alpha_i\tilde{\bf g}_i\tilde{\bf g}_i^{H}+\sigma^2{\bf I} \right)^{-1}\tilde{\bf g}_k\geq C_1, \nonumber 
 \end{align}
\begin{align}
\text{and} \ \ \frac{1}{K\sigma^2}\tilde{\bf g}_k^{H}\tilde{\bf g}_k\leq C_2. 
\end{align}
 Hence, for any feasible $\tau$, we have:
 $$
 \tau\leq C_2 \alpha_0\ \ \text{and} \ \ \tau\geq \alpha_0C_1, 
 $$
which proves that $\tau$ lies almost surely in a compact set of $(0,\infty)$. Let $\mathcal{I}$ be a compact interval in $(0,\infty)$ containing $(C_1\alpha_0,C_2\alpha_0)$. We may thus exploit Theorem \ref{th:1} to show that:
$$
\sup_{\tau\in\mathcal{I}} \max_{1\leq k\leq K}\left|\frac{\tau}{d_k(\tau)}-\frac{\tau}{\overline{d}(\tau)}\right|\underset{M \rightarrow \infty}{\longrightarrow} 0.
$$
Hence, for any $\delta>0$, there exists $N_0, M_0$ and $K_0$ such that for all $N\geq N_0$, $M\geq M_0$ and $K\geq K_0$, we have:
\begin{equation}
\forall \tau\in\mathcal{I}\ \ \forall k \in \mathcal{K} \ \ \frac{\tau}{\overline{d}(\tau)}-\delta \leq \frac{\tau}{d_k(\tau)} \leq \frac{\tau}{\overline{d}(\tau)}+\delta.  \label{eq:res}
\end{equation}
Now, consider the following perturbed asymptotically equivalent problems:
\begin{equation}
\tag{$\mathcal{P}_{\rm asym}^{+}$}
\begin{aligned}
&	\tilde{\tau}_{\delta^{+}}= 	\displaystyle{\max_{{d},\tau}}  \ \  \tau  \\
 &  \text { s.t. }   \ \ {d}=\frac{1}{K}\text{Trace} \left \lbrace  \vec{U}\vec{U}^H \left(\vec{U}\vec{U}^H  \frac{\tau}{{{d}}(1+\tau)}+ {\sigma}^2 \mathbf{I}_{M} \right)^{-1} \right \rbrace\\
&  \ \ \frac{\tau}{{d}}\leq \alpha_k+\delta , \ \ \forall k \in \mathcal{K},
\end{aligned}
\end{equation}
and 
\begin{equation}
\tag{$\mathcal{P}_{\rm asym}^{-}$}
\begin{aligned}
&	\tilde{\tau}_{\delta^{-}}= 	\displaystyle{\max_{{d},\tau}}     \tau  \\
 & \text { s.t. }  {d}=\frac{1}{K}\text{Trace} \left \lbrace  \vec{U}\vec{U}^H \left(\vec{U}\vec{U}^H  \frac{\tau}{{{d}}(1+\tau)}+ {\sigma}^2 \mathbf{I}_{M} \right)^{-1} \right \rbrace\\
& \frac{\tau}{{d}}\leq \alpha_k-\delta , \ \ \forall k \in \mathcal{K}.
\end{aligned}
\end{equation}
The main idea is to show that almost surely, for sufficiently large dimensions:
\begin{equation}
\tilde{\tau}_{\delta^{-}}\leq \tau^* \leq \tilde{\tau}_{\delta^{+}}. \label{eq:trt}
\end{equation}
Then, based on Lemma 9 in \cite{kammoun2021precise}, we have
$$
\lim_{\delta\to 0}\tilde{\tau}_{\delta^{-}}=\tilde{\tau},
$$
and 
$$
\lim_{\delta\to 0}\tilde{\tau}_{\delta^{+}}=\tilde{\tau}.
$$
Hence, for any $\epsilon>0$, we may choose $\delta$ sufficiently small such that:
\begin{equation}
\tilde{\tau}\geq \tilde{\tau}_{\delta^{+}}-\epsilon \ \ \text{and} \ \ \tilde{\tau}\leq \tilde{\tau}_{\delta_{-}}+\epsilon. \label{eq:trtr}
\end{equation}
By combining this with \eqref{eq:trt}, we thus obtain
$$
\tilde{\tau}-\epsilon\leq \tau^* \leq \tilde{\tau}+\epsilon.
$$
which shows the desired. 
We will thus focus in the sequel on establishing both inequalities in \eqref{eq:trt}.\\
\underline{\bf Proof of $\tau^*\leq \tilde{\tau}_{\delta^+}$.} Let $\tau$ be feasible for Problem $\hat{\mathcal{P}}$. To prove the desired, it suffices to show that $\tau$ is feasible for Problem $\mathcal{P}_{\rm asym}^{+}$. Indeed, it follows from \eqref{eq:res} that:
$$
\frac{\tau}{\overline{d}(\tau)}-\delta\leq\frac{\tau}{d_k(\tau)}\leq \alpha_k,
$$
and hence
$$
\frac{\tau}{\overline{d}(\tau)}\leq \alpha_k+\delta. 
$$
This proves that $\tau$ is feasible for  $\mathcal{P}_{\rm asym}^{+}$ and thus $\tau^\star\leq \tilde{\tau}_{\delta^+}$.\\
\underline{\bf Proof of $\tau^*\geq \tilde{\tau}_{\delta^-}$.} Let $\tau$ be feasible for $\mathcal{P}_{\rm asym}^{-}$. Then, $\frac{\tau}{\overline{d}(\tau)}\leq \alpha_k-\delta$. Using \eqref{eq:res}, we thus get:
$$
\frac{\tau}{\overline{d}_k(\tau)}-\delta\leq \alpha_k-\delta,
$$
or equivalently
$$
\frac{\tau}{d_k(\tau)}\leq \alpha_k.
$$
Hence $\tau$ is feasible for $\hat{\mathcal{P}}$ and thus $\tau^\star\geq \tilde{\tau}_{\delta}^{-}$. 
\\
\noindent{\bf Second part.} 
First, we check that \eqref{eq:tau_bar} admits a unique solution. Indeed, by dividing both sides of \eqref{eq:tau_bar}, we obtain:
\begin{equation}
1=\frac{\alpha_0}{K}\text{Trace}\left\{{\bf U}{\bf U}^{H}\left(\frac{{\bf U}{\bf U}^{H}\alpha_0\tau}{1+\tau}+\sigma^2\tau{\bf I}\right)^{-1}\right\}. \label{eq:tt}
\end{equation}
The right-hand side of the above equality is a decreasing function of $\tau$ tending to $0$ as $\tau\to\infty$ and to $\infty$ as $\tau$ goes to zero. Hence, there exists a unique $\overline{\tau}$ satisfying \eqref{eq:tt}. Let $\overline{d}=\frac{\overline{\tau}}{\alpha_0}$ where we recall that $\alpha_0=\min_{1\leq k\leq K}\alpha_k$. Our aim is to show that $\tilde{\tau}=\overline{\tau}$ and $\overline{d}=\breve{d}$. For that, it suffices to show that $\frac{\tilde{\tau}}{\breve{d}}=\alpha_0$. Indeed, if this is true, then starting from
$$
\breve{d}=\frac{1}{K}\text{Trace} \left \lbrace  \vec{U}\vec{U}^H \left(\vec{U}\vec{U}^H  \frac{\tilde{\tau}}{{\breve{d}}(1+\tilde{\tau})}+ {\sigma}^2 \mathbf{I}_{M} \right)^{-1} \right \rbrace,
$$
we thus obtain:
$$
\frac{\breve{d}}{\tilde{\tau}}=\frac{1}{K}\text{Trace} \left \lbrace  \vec{U}\vec{U}^H \left(\vec{U}\vec{U}^H  \frac{\tilde{\tau}^2}{{\breve{d}}(1+\tilde{\tau})}+ \tilde{\tau}{\sigma}^2 \mathbf{I}_{M} \right)^{-1} \right \rbrace.
$$
Replacing $\frac{\tilde{\tau}}{\breve{d}}$ by $\alpha_0$, we thus obtain:
$$
1=\frac{\alpha_0}{K}\text{Trace}\left\{{\bf U}{\bf U}^{H}\left( {\bf U}{\bf U}^{H} \alpha_0\frac{\tilde{\tau}}{1+\tilde{\tau}}+\tilde{\tau}\sigma^2{\bf I}\right)^{-1}\right\},
$$
or equivalently
$$
\tilde{\tau}=\frac{\alpha_0}{K}\text{Trace}\left\{{\bf U}{\bf U}^{H}\left( {\bf U}{\bf U}^{H} \alpha_0\frac{1}{1+\tilde{\tau}}+\sigma^2{\bf I}\right)^{-1}\right\}. 
$$
Since $\overline{\tau}$ is the unique solution to the above equation, we thus have $\tilde{\tau}=\overline{\tau}$, and $\overline{d}=\breve{d}$. 
In the sequel, we will thus focus on showing that $\frac{\tilde{\tau}}{\breve{d}}=\alpha_0$. We will thus proceed by contradiction and assume that $\frac{\tilde{\tau}}{\breve{d}}<\alpha_0$. Then, starting from
$$
\breve{d}=\frac{1}{K}\text{Trace}\left \lbrace {\bf U}{\bf U}^{H}\left(\frac{\tilde{\tau}}{\breve{d}(1+\tilde{\tau})}{\bf U}{\bf U}^{H}+\sigma^2{\bf I}\right)^{-1}\right \rbrace , 
$$
we have
\begin{align}
\breve{d}&> \frac{1}{K}\text{Trace}\left \lbrace {\bf U}{\bf U}^{H}\left(\frac{\alpha_0}{(1+\tilde{\tau})}{\bf U}{\bf U}^{H}+\sigma^2{\bf I}\right)^{-1}\right \rbrace .
\end{align}
Next, we use the fact that $\tilde{\tau}\geq \overline{\tau}$ to obtain
\begin{align}
\breve{d}&> \frac{1}{K}\text{Trace}\left \lbrace {\bf U}{\bf U}^{H}\left(\frac{\alpha_0}{1+\overline{\tau}}{\bf U}{\bf U}^{H}+\sigma^2{\bf I}\right)^{-1}\right \rbrace, \\
&= \frac{1}{K}\text{Trace}\left \lbrace {\bf U}{\bf U}^{H}\left(\frac{\overline{\tau}}{(1+\overline{\tau})\overline{d}}{\bf U}{\bf U}^{H}+\sigma^2{\bf I}\right)^{-1} \right \rbrace ,
\end{align}
or equivalently $\breve{d}>\overline{d}$. 
To continue, we start from 
$$
\breve{d}=\frac{1}{K}\text{Trace} \left \lbrace  \vec{U}\vec{U}^H \left(\vec{U}\vec{U}^H  \frac{\tilde{\tau}}{{\breve{d}}(1+\tilde{\tau})}+ {\sigma}^2 \mathbf{I}_{M} \right)^{-1} \right \rbrace, 
$$
and use the fact that $\tilde{\tau}\geq \overline{\tau}$ to obtain:
$$
\breve{d}\leq \frac{1}{K}\text{Trace} \left \lbrace  \vec{U}\vec{U}^H \left(\vec{U}\vec{U}^H  \frac{\overline{\tau}}{{\breve{d}}(1+\overline{\tau})}+ {\sigma}^2 \mathbf{I}_{M} \right)^{-1} \right \rbrace, 
$$
or equivalently
$$
\breve{d}\leq \frac{\overline{d}}{K}\text{Trace}\left \lbrace {\bf U}{\bf U}^{H}\left({\bf U}{\bf U}^{H}\frac{\overline{\tau}}{1+\overline{\tau}}+\sigma^2\overline{d}{\bf I}\right)^{-1} \right \rbrace. 
$$
Now, since $\breve{d}>\overline{d}$, we obtain
\begin{align}
\breve{d}&< \frac{\overline{d}}{K}\text{Trace}\left \lbrace {\bf U}{\bf U}^{H}\left({\bf U}{\bf U}^{H}\frac{\overline{\tau}}{1+\overline{\tau}}+\sigma^2\breve{d}{\bf I}\right)^{-1} \right \rbrace, \\
&\leq \frac{\overline{d}}{K}\text{Trace}\left \lbrace {\bf U}{\bf U}^{H}\left({\bf U}{\bf U}^{H}\frac{\overline{\tau}}{1+\overline{\tau}}+\sigma^2\overline{d}{\bf I}\right)^{-1} \right \rbrace , 
\end{align}
or equivalently $\breve{d}<\overline{d}$, which is in contradiction with the previously obtained inequality $\breve{d}>\overline{d}$.

\end{appendices}

\bibliographystyle{IEEEtran}
\bibliography{reference_asymp}

\vfill

\end{document}